\documentstyle [12pt,amsfonts] {article}
\input epsf

\newcommand{\beq}{\begin{equation}}
\newcommand{\eeq}{\end{equation}}
\newcommand{\beqs}{\begin{eqnarray}}
\newcommand{\eeqs}{\end{eqnarray}}

\catcode`@=11
\@addtoreset{equation}{section}
\@addtoreset{equation}{subsection}
\def\theequation{\ifnum\value{section}=0 \arabic{equation}\ignorespaces
\else \ifnum\value{section}=-1 A.\arabic{equation}\ignorespaces
\else \ifnum\value{subsection}=0 \thesection.\arabic{equation}\ignorespaces
\else \thesection.\arabic{subsection}.\arabic{equation}\ignorespaces
                           \fi
                      \fi
                 \fi}
\catcode`@=12

\begin{document}

\def\thefootnote{\fnsymbol{footnote}}

\baselineskip 6.0mm

\vspace{4mm}

\begin{center}

{\Large \bf $T=0$ Partition Functions for Potts Antiferromagnets on 
Lattice Strips with Fully Periodic Boundary Conditions} 

\vspace{8mm}

\setcounter{footnote}{0}
Shu-Chiuan Chang$^{(a)}$\footnote{email: shu-chiuan.chang@sunysb.edu} and
\setcounter{footnote}{6}
Robert Shrock$^{(a,b)}$\footnote{(a): permanent address;
email: robert.shrock@sunysb.edu}

\vspace{6mm}

(a) \ C. N. Yang Institute for Theoretical Physics  \\
State University of New York       \\
Stony Brook, N. Y. 11794-3840  \\

(b) \ Physics Department \\
Brookhaven National Laboratory \\
Upton, NY  11973

\vspace{10mm}

{\bf Abstract}
\end{center}

We present exact calculations of the zero-temperature partition function for
the $q$-state Potts antiferromagnet (equivalently, the chromatic polynomial)
for families of arbitrarily long strip graphs of the square and triangular
lattices with width $L_y=4$ and boundary conditions that are doubly periodic or
doubly periodic with reversed orientation (i.e. of torus or Klein bottle type).
These boundary conditions have the advantage of removing edge effects. In the
limit of infinite length, we calculate the exponent of the entropy, $W(q)$ and
determine the continuous locus ${\cal B}$ where it is singular.  We also give
results for toroidal strips involving ``crossing subgraphs''; these make
possible a unified treatment of torus and Klein bottle boundary conditions and
enable us to prove that for a given strip, the locus ${\cal B}$ is the same for
these boundary conditions.

\vspace{16mm}

\pagestyle{empty}
\newpage

\pagestyle{plain}
\pagenumbering{arabic}
\renewcommand{\thefootnote}{\arabic{footnote}}
\setcounter{footnote}{0}

\section{Introduction}

The $q$-state Potts antiferromagnet (AF) \cite{potts,wurev} exhibits nonzero
ground state entropy, $S_0 > 0$ (without frustration) for sufficiently large
$q$ on a given lattice $\Lambda$ or, more generally, on a graph $G=(V,E)$
defined by its set of vertices $V$ and edges joining these vertices $E$. This 
is
equivalent to a ground state degeneracy per site $W > 1$, since $S_0 = k_B \ln
W$.  Such nonzero ground state entropy is important as an exception to the
third law of thermodynamics \cite{cw}.  There is a close connection with
graph theory
here, since the zero-temperature partition function of the above-mentioned
$q$-state Potts antiferromagnet on a graph $G$ satisfies
\beq
Z(G,q,T=0)_{PAF}=P(G,q)
\label{zp}
\eeq
where $P(G,q)$ is the chromatic polynomial expressing the number of ways of
coloring the vertices of the graph $G$ with $q$ colors such that no two
adjacent vertices have the same color (for reviews, see
\cite{rrev}-\cite{bbook}).  The minimum number of colors necessary for such a
coloring of $G$ is called the chromatic number, $\chi(G)$.
Thus
\beq
W(\{G\},q) = \lim_{n \to \infty} P(G,q)^{1/n}
\label{w}
\eeq
where $n=|V|$ is the number of vertices of $G$ and $\{G\} = \lim_{n \to
\infty}G$.  At certain special
points $q_s$ (typically $q_s=0,1,.., \chi(G)$), one has the noncommutativity of
limits
\beq
\lim_{q \to q_s} \lim_{n \to \infty} P(G,q)^{1/n} \ne \lim_{n \to
\infty} \lim_{q \to q_s}P(G,q)^{1/n}
\label{wnoncom}
\eeq
and hence it is necessary to specify the
order of the limits in the definition of $W(\{G\},q_s)$ \cite{w}. Denoting
$W_{qn}$ and $W_{nq}$ as the functions defined by the different order of limits
on the left and right-hand sides of (\ref{wnoncom}), we take $W \equiv W_{qn}$
here; this has the advantage of removing certain isolated discontinuities that
are present in $W_{nq}$.

Using the expression for $P(G,q)$, one can generalize $q$ from ${\mathbb Z}_+$
to ${\mathbb C}$.  The zeros of $P(G,q)$ in the complex $q$ plane are called
chromatic zeros; a subset of these may form an accumulation set in the $n \to
\infty$ limit, denoted ${\cal B}$ \cite{bkw}, which is the continuous locus of
points where $W(\{G\},q)$ is nonanalytic.~\footnote{\footnotesize{For some
families of graphs ${\cal B}$ may be null, and $W$ may also be nonanalytic at
certain discrete points.}}  The maximal region in the complex $q$ plane to
which one can analytically continue the function $W(\{G\},q)$ from physical
values where there is nonzero ground state entropy is denoted $R_1$.  The
maximal value of $q$ where ${\cal B}$ intersects the (positive) real axis is
labelled $q_c(\{G\})$.  This point is important since it separates the interval
$q > q_c(\{G\})$ on the positive real $q$ axis where the Potts model (with $q$
extended from ${\mathbb Z}_+$ to ${\mathbb R}$) exhibits nonzero ground state
entropy (which increases with $q$, asymptotically approaching $S_0 = k_B \ln q$
for large $q$, and which for a regular lattice $\Lambda$ can be calculated
approximately via large--$q$ series expansions) from the interval $0 \le q \le
q_c(\{G\})$ in which $S_0$ has a different analytic form.  Early calculations
of chromatic polynomials for $L_y=2$ strips of the square lattice with periodic
longitudinal boundary conditions were performed in \cite{bds} (see also the
related works \cite{bm}-\cite{read91}).

Here we present exact calculations of the chromatic polynomials for strips of
the square and triangular lattice with transverse width $L_y=4$
(i.e. transverse cross sections forming squares) and arbitrarily great length
$L_x$ with the following boundary conditions: (i) $(PBC_y,PBC_x)=$ toroidal,
and (ii) $(PBC_y,TPBC_x)=$ Klein bottle, where $PBC_i$ denotes periodic
boundary conditions in the $i$'th direction and $TPBC_x$ denotes periodic
longitudinal boundary conditions with an orientation-reversal
(twist).\footnote{\footnotesize{The boundary conditions $(PBC_y,PBC_x)$ and
$(PBC_y,TPBC_x)$ can be implemented in a manner that is uniform in the length
$L_x$; as noted before \cite{tk}, the boundary conditions $(TPBC_y,PBC_x)$
(different type of Klein bottle) and $(TPBC_y,TPBC_x)$ (projective plane)
require different identifications as $L_x$ varies and will not be considered
here.}}  These extend our previous calculations of chromatic polynomials for
width $L_y=3$ on the square \cite{tk} and triangular \cite{t} lattices with
torus and Klein bottle boundary conditions.

A major motivation for using boundary conditions that are fully periodic
or fully periodic with reversed orientation (here, toroidal and Klein bottle)
is the well-known fact that if one imposes periodic boundary conditions in a
certain direction, this removes edge effects in that direction.  Clearly the
most complete removal of such edge effects is achieved if one imposes fully
periodic boundary conditions (including the possibility of orientation
reversal).  This also has an important related consequence pertaining to the
uniformity of the lattice.  To discuss this, we first recall two definitions
from mathematical graph theory.  The degree $\Delta$ of a vertex of a graph is
the number of edges connected to it.  A $\Delta$-regular graph is a graph in
which all vertices have the same degree, $\Delta$.  An infinite regular lattice
has the property that each vertex (site) on the lattice has the same degree,
i.e., coordination number.  For the two types of lattices considered here,
namely square and triangular, the coordination number is 4 and 6,
respectively.  It is advantageous to deal with finite sections of
regular lattices having boundary conditions that preserve the $\Delta$-regular
property of the infinite lattice.  Fully periodic periodic boundary conditions,
and the reversed-orientation periodic boundary conditions considered here, have
the merit of preserving this property of $\Delta$-regularity; in contrast, this
is not the case if one uses boundary conditions that are free in one or more
directions. In previous studies with families of lattice strip graphs of
arbitrarily great length with periodic or reversed-orientation periodic
longitudinal boundary conditions and free transverse boundary conditions (i.e.,
cyclic or M\"obius strips), it was shown that, in the $L_x \to \infty$ limit,
the resultant locus ${\cal B}$ exhibits, for finite width $L_y$, a number of
properties expected to hold for the locus ${\cal B}$ on the infinite 2D
lattice, including (i) passing through $q=0$, (ii) passing through $q=2$, (iii)
passing through a maximal real point, $q_c$, and (iv) enclosing one or more
regions including the interval $0 < q < q_c$ \cite{w}, \cite{pg}-\cite{bcc}. 
In contrast, if one uses free longitudinal boundary conditions, it was found in
\cite{strip}-\cite{hs} that properties (i) and (iv) do not hold, and 
properties (ii) and (iii) do not, in general, hold; rather, one anticipates
that these would be approached in the limit $L_y \to \infty$.  It
was thus inferred that the key condition to guarantee that these properties
hold is the presence of periodic (or reversed-orientation periodic)
longitudinal boundary conditions \cite{bcc}.  This thus provides a third
motivation for calculations with doubly periodic boundary conditions, since one
expects that the resultant loci ${\cal B}$ will exhibit the features (i)-(iv)
already for finite $L_y$, and this was confirmed by the study of $L_y=3$ strips
of the square \cite{tk} and triangular \cite{t} lattices.  As will be seen, our
exact results for $L_y=4$ again support this inference.  A fourth motivation
for this study is that, as was shown in the earlier calculations of chromatic
polynomials for strips of the square \cite{tk} and triangular lattices \cite{t}
with width $L_y=3$ and is again true for width $L_y=4$, the use of Klein
bottle, as opposed to torus, boundary conditions has the effect of simplifying
the structure of the resultant chromatic polynomial.  This thus elucidates the
effect of the topology of the surface on which the family of strip graphs is
embedded with the structure of the chromatic polynomial.  In addition to those
listed, some previous related calculations of chromatic polynomials for
families of graphs with periodic longitudinal boundary conditions are in
Refs. \cite{bds}-\cite{cf}.

In general, the $L_y \times L_x$ strips of the square and triangular lattice
have $n=|V|=L_yL_x$ vertices and, for the number of edges $|E|=(\Delta/2)n$ the
values $|E|=2n$ and $|E|=3n$ respectively. (For $L_x=2$, some of these strip
graphs involve multiple edges joining pairs of vertices and hence are
multigraphs rather than proper graphs; we shall be interested primarily in the
cases $L_x \ge 3$ where there are no multiple edges.)
 
We label a particular type of strip graph as $G_s$ or just $G$ and the 
specific graph of width $L_y$ and length $L_x$ vertices as $(G_s,L_y \times
L_x,BC_y,BC_x)$. 
A generic form for chromatic polynomials for recursively defined families of 
graphs, of which strip graphs $G_s$ are special cases, is \cite{bkw} 
\beq
P(G_s,L_y \times L_x,BC_y,BC_x,q) =  
\sum_{j=1}^{N_{G_s,\lambda}} c_{G_s,j}(q)(\lambda_{G_s,j}(q))^m
\label{pgsum}
\eeq 
where $c_{G_s,j}(q)$ and the $N_{G_s,\lambda}$ terms $\lambda_{G_s,j}(q)$ 
depend on the type of strip graph $G_s$ but are independent of $m$.  The
$\lambda_{G_s,j}$ are the (nonzero) eigenvalues of the coloring matrix
\cite{b,matmeth,cf}.  We shall denote the total number of different eigenvalues
of the coloring matrix for a recursive family of graphs $G_s$ as
$N_{G_s,\lambda,tot}$.  Clearly $N_{G_s,\lambda,tot}=N_{G_s,\lambda}$ if
there is no zero eigenvalue, and $N_{G_s,\lambda,tot}=N_{G_s,\lambda}+1$ if
there is a zero eigenvalue.  Our results illustrate both of these
possibilities.

For a given type of strip graph $G_s$, we denote the sum of the coefficients 
$c_{G_s,j}$ as 
\beq
C_{G_s} \equiv C(G_s)=\sum_{j=1}^{N_{G_s,\lambda}} c_{G_s,j} \ . 
\label{cgsum}
\eeq
According to a general theorem, for a strip $G_s$ of the square or
triangular
lattice with torus boundary conditions \cite{pm,cf}, 
\beq
C(G_s, L_y \times L_x,PBC_y,PBC_x)=P(C_{L_y},q) \ , \quad G_s=sq,tri
\label{csumtor}
\eeq
where $C_n$ denotes the circuit graph with $n$ vertices and 
$P(C_n,q)=(q-1)^n+(q-1)(-1)^n$. Further, for a strip of the square or
triangular lattice with Klein bottle boundary conditions \cite{cf}
\beq
C(G_s, L_y \times L_x,PBC_y,TPBC_x)=0 \ , \quad G_s=sq,tri \ . 
\label{csumkb}
\eeq

\section{$L_y=4$ Strip of the Square Lattice with $(PBC_y,PBC_x)$}

In general, for a strip of the square lattice of size $L_y \times L_x$ with
$(PBC_y,PBC_x)$, i.e., toroidal boundary conditions,
for $L_y \ge 2$ and $L_x \ge 2$, the chromatic number is given by
\beq
\chi(sq,L_y \times L_x,PBC_y,PBC_x) = \cases{ 2 & if $L_y$ is even and 
$L_x$ is even \cr
        3 & otherwise}
\label{chisqtorus}
\eeq
Thus, in the present case with $L_y=4$, it follows that 
$\chi=2$ for even $L_x$ and $\chi=3$ for odd $L_x$. 
We calculate the chromatic polynomial $P$ by a systematic, iterative use of 
the deletion-contraction theorems as in our earlier work \cite{strip,hs} and
a coloring matrix method \cite{b}.  
For the $L_y=4$ strip graphs of the square lattice with
torus boundary conditions (labelled $st4$), we find 
$N_{st4,\lambda}=33$ and 
\beq
P(sq, 4 \times L_x,PBC_y,PBC_x,q) = \sum_{j=1}^{33} c_{st4,j} 
(\lambda_{st4,j})^{L_x}
\label{psqtorus}
\eeq
where
\beq
\lambda_{st4,1}=1 
\label{lamtorsq1}
\eeq
\beq
\lambda_{st4,2}=1-q
\label{lamtorsq2}
\eeq
\beq
\lambda_{st4,3}=2-q
\label{lamtorsq3}
\eeq
\beq
\lambda_{st4,4}=3-q
\label{lamtorsq4}
\eeq
\beq
\lambda_{st4,5}=4-q
\label{lamtorsq5}
\eeq
\beq
\lambda_{st4,6}=5-q
\label{lamtorsq6}
\eeq
\beq
\lambda_{st4,7}=q^2-5q+5
\label{lamtorsq7}
\eeq
\beq
\lambda_{st4,8}=q^2-5q+7
\label{lamtorsq8}
\eeq
\beq
\lambda_{st4,9}=(q-1)(q-3)
\label{lamtorsq9}
\eeq
\beqs
& & \lambda_{st4,(10,11)}=\frac{1}{2}\biggl [ q^4-8q^3+29q^2-55q+46 \cr\cr
& & \pm \Bigl (
q^8-16q^7+118q^6-526q^5+1569q^4-3250q^3+4617q^2-4136q+1776 \Bigr )^{1/2} \
\biggr ] \cr\cr
& & 
\label{lamtorsq1011}
\eeqs
\beqs
& & \lambda_{st4,(12,13)}=\frac{1}{2}\biggl [-(q^3-7q^2+18q-17) \cr\cr 
& & \pm \Bigl ( q^6-14q^5+81q^4-250q^3+442q^2-436q+193 \Bigr )^{1/2} \
\biggr ]
\label{lamtorsq1213}
\eeqs
\beq
\lambda_{st4,(14,15)}=\frac{1}{2}\biggl [ q^2-7q+9 \pm \Bigl (
q^4-10q^3+35q^2-50q+33 \Bigr )^{1/2} \ \biggr ] 
\label{lamtorsq1415}
\eeq
\beq
\lambda_{st4,(16,17)}=\frac{1}{2}\biggl [ (q-3)^2 \pm \Bigl (
q^4-8q^3+26q^2-48q+41 \Bigr )^{1/2} \ \biggr ] 
\label{lamtorsq1617}
\eeq
\beq
\lambda_{st4,(18,19)}=\frac{1}{2}\biggl [ q^2-6q+11 \pm (q-3)\Bigl (
q^2-2q+9 \Bigr )^{1/2} \ \biggr ]
\label{lamtorsq1819}
\eeq
\beq
\lambda_{st4,(20,21)}=3-q \pm \sqrt{3} \ . 
\label{lamtorsq2021}
\eeq
The remaining twelve $\lambda_{st4,j}$'s for $22 \le j \le 33$ are roots of 
four cubic equations, 
\beqs
& & \xi^3+(q^3-6q^2+16q-14)\xi^2-(q-1)(q^4-9q^3+31q^2-55q+43)\xi \cr\cr
& & -(q-3)(q-1)^2(q^3-6q^2+12q-10)=0
\label{eqcub1sq}
\eeqs
with roots $\lambda_{st4,j}$ for $j=22,23,24$, 
\beqs
& & \xi^3+(q-4)(q^2-6q+12)\xi^2-(q-3)(q^4-11q^3+45q^2-81q+59)\xi \cr\cr
& & -(q^6-15q^5+91q^4-285q^3+488q^2-442q+170)=0
\label{eqcub2sq}
\eeqs
with roots $\lambda_{st4,j}$ for $j=25,26,27$, 
\beqs
& & \xi^3-2(q^2-6q+12)\xi^2+ (q^4-13q^3+59q^2-113q+83)\xi \cr\cr
& & +(q^5-13q^4+62q^3-135q^2+141q-60)=0
\label{eqcub3sq}
\eeqs
with roots $\lambda_{st4,j}$ for $j=28,29,30$, and 
\beqs
& & \xi^3-2(q^2-6q+10)\xi^2+(q^4-13q^3+59q^2-113q+75)\xi \cr\cr
& & +(q^5-13q^4+64q^3-149q^2+167q-72)=0
\label{eqcub4sq}
\eeqs
with roots $\lambda_{st4,j}$ for $j=31,32,33$. 

The corresponding coefficients are
\beq
c_{st4,1} = q^4-8q^3+20q^2-15q+1
\label{cst4_1}
\eeq
\beq
c_{st4,2} = \frac{1}{2}c_{st4,4}=c_{st4,6}=\frac{1}{3}(q-1)(q^2-5q+3)
\label{cst4_2}
\eeq
\beq
c_{st4,3} = \frac{1}{6}(q-2)(4q^2-13q-3)
\label{cst4_3}
\eeq
\beq
c_{st4,j}=\frac{2}{3}q(q-2)(q-4) \quad {\rm for} \ \ j=5, 20, 21
\label{cst4_5}
\eeq
\beq
c_{st4,j}=\frac{1}{2}(q-1)(q-2) \quad {\rm for} \ \ j=7, 18, 19
\label{cst4_7}
\eeq
\beq
c_{st4,j}=(q-1)(q-2) \quad {\rm for} \ \ j=31,32,33
\label{cst43133}
\eeq
\beq
c_{st4,j}=\frac{1}{2}q(q-3) \quad {\rm for} \ \ j=8,14,15,28,29,30
\label{cst4_8}
\eeq
\beq
c_{st4,j}=q(q-3)  \quad {\rm for} \ \ j=16,17
\label{cst1617}
\eeq
\beq
c_{st4,j}=1 \quad {\rm for} \ \ j=9,10,11
\label{cst4_911}
\eeq
\beq
c_{st4,j}=2(q-1) \quad {\rm for} \ \ j=12,13
\label{ctorsq1213}
\eeq
\beq
c_{st4,j}=q-1  \quad {\rm for} \quad 22 \le j \le 27 \ . 
\label{ctorsq2227}
\eeq
The sum of these coefficients is equal to $P(C_4,q)=q(q-1)(q^2-3q+3)$, as 
dictated by the $L_y=4$ special case of our general result (\ref{csumtor}).

\begin{figure}[hbtp]
\centering
\leavevmode
\epsfxsize=3.0in
\begin{center}
\leavevmode
\epsffile{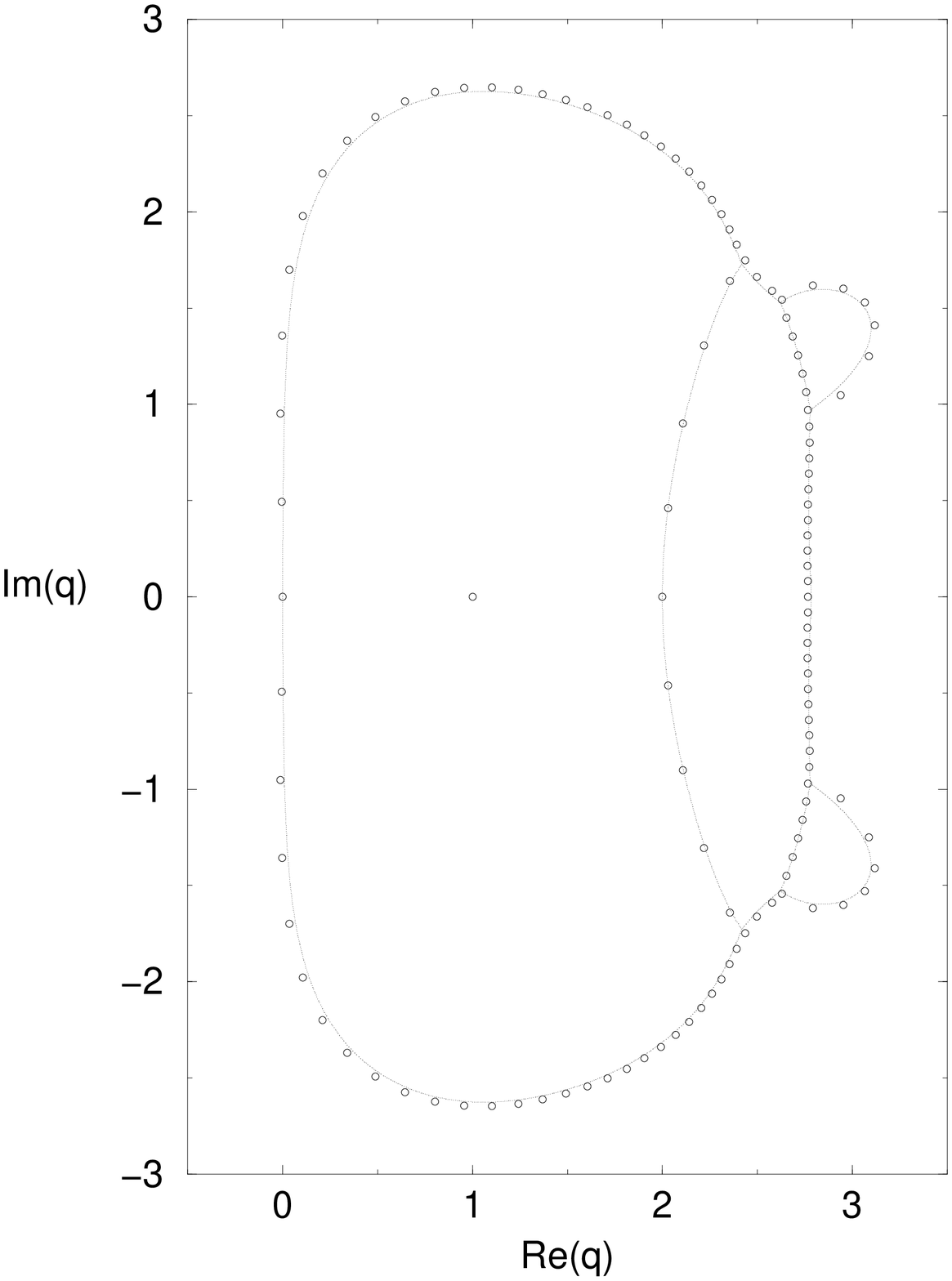}
\end{center}
\caption{\footnotesize{Singular locus ${\cal B}$ for the $L_x \to \infty$ limit
of the strip of the square lattice with $L_y=4$ and toroidal boundary
conditions. For comparison, chromatic zeros are shown for $L_x=30$ (i.e., 
$n=120$).}}
\label{sqpxpy4}
\end{figure}

The singular locus ${\cal B}$ for the $L_x \to \infty$ limit of the strip of
the square lattice with $L_y=4$ and toroidal boundary conditions is shown in
Fig. \ref{sqpxpy4}.  For comparison, chromatic zeros are calculated and shown 
for length $L_x=30$ (i.e., $n=120$ vertices). 
The locus ${\cal B}$ crosses the real axis at the points $q=0$, $q=2$, and at 
the maximal point $q=q_c$, where
\beq
q_c=2.7827657...  \quad {\rm for} \ \ \{G\}=(sq, 4 \times \infty,PBC_y,PBC_x) .
\label{qcsq}
\eeq
As is evident from Fig. \ref{sqpxpy4}, the locus ${\cal B}$ separates the $q$ 
plane into different regions including the following: 
(i) $R_1$, containing the semi-infinite intervals $q > q_c$ and $q < 0$ on the
real axis and extending outward to infinite $|q|$; (ii) $R_2$ containing the
interval $2 < q < q_c$; (iii) $R_3$ containing the real interval $0 < q < 2$;
and (iv) the complex-conjugate pair $R_4,R_4^*$ centered approximately at 
$q=2.9 \pm 1.3i$.  The (nonzero) density of chromatic zeros has the smallest
values on the curve separating regions $R_1$ and $R_3$ in the vicinity of the
point $q=0$ and on the curve separating regions $R_2$ and $R_3$ in the 
vicinity of the point $q=2$. 

In region $R_1$, $\lambda_{st4,10}$ is the dominant $\lambda_{G,j}$, so 
\beq
W = (\lambda_{st4,10})^{1/4} \ , \quad q \in R_1 \ . 
\label{wsqr1}
\eeq
This is the same as $W$ for the corresponding $L_x \to \infty$ limit of the 
strip of the square lattice with the same width $L_y=4$ and cylindrical 
$(PBC_y,FBC_x)$ boundary conditions, calculated in \cite{strip2}.  
This equality of 
the $W$ functions for the $L_x \to \infty$ limit of two strips of a given
lattice with the same transverse boundary conditions and different longitudinal
boundary conditions in the more restrictive region $R_1$ defined by the two
boundary conditions is a general result \cite{wcy,bcc}. 

In region $R_2$, the largest root of the cubic equation (\ref{eqcub3sq}) is
dominant; we label this as $\lambda_{st4,28}$ so that 
\beq
|W| = |\lambda_{st4,28}|^{1/4} \ , \quad q \in R_2
\label{wsqr2}
\eeq
(in regions other than $R_1$, only $|W|$ can be determined unambiguously 
\cite{w}). Thus, $q_c$ is the relevant solution of the equation of degeneracy
in magnitude $|\lambda_{st4,10}|=|\lambda_{st4,28}|$.  In region $R_3$,
\beq
|W|=|\lambda_{st4,25}|^{1/4} \ , \quad q \in R_3 \ . 
\label{wsqr3}
\eeq 
In regions $R_4,R_4^*$, 
\beq
|W|=|\lambda_{st4,22}|^{1/4} \ , \quad q \in R_4, \ R_4^* \ .
\label{wsqr4}
\eeq

In \cite{s4} we have listed values of $W$ for a range of values of $q$ for the
$L_x \to \infty$ limit of various strips of the square lattice, including $(sq,
4 \times \infty,PBC_y,FBC_x)$.  Since $W$ is independent of $BC_x$ for $q$ in
the more restrictive region $R_1$ defined by $FBC_x$ and $(T)PBC_x$ (which is
the $R_1$ defined by $PBC_x$ here), it follows, in particular, that 
\beq
W(4 \times \infty,PBC_y,(T)PBC_x,q)=W(4 \times \infty,PBC_y,FBC_x,q) \quad 
{\rm for} \ \ q \ge q_c
\label{wsqtor4qgeqc}
\eeq
where $q_c$ was given above in (\ref{qcsq}). 
For low integral values of $q$ we list the values of $|W(q)|$ for this strip in
Table \ref{wsqinternal}, together with corresponding values given in 
\cite{s4} for $W$ in the $L_x \to \infty$ limit of the $L_y=3$ strip with
$(PBC_y,(T)PBC_x)$. 

\begin{table}
\caption{\footnotesize{Values of $W(sq,L_y \times \infty,PBC_y,(T)PBC_x,q)$ for
low integral $q$ and for respective $q_c$.}}
\begin{center}
\begin{tabular}{|c|c|c|c|c|c|c|c|}
\hline\hline
$L_y$ & $BC_y$ & $BC_x$ & $|W_{q=0}|$ & $|W_{q=1}|$ & $|W_{q=2}|$ &
$q_c$ & $W_{q=q_c}$ \\
\hline\hline
3  & P & (T)P & 2.35   & 1.91    & 1.44   & 3      & 1.26   \\ \hline 
4  & P & (T)P & 2.58   & 2.11    & 1.64   & 2.78   & 1.44   \\ \hline\hline
\end{tabular}
\end{center}
\label{wsqinternal}
\end{table}

For various lengths $L_x$, some of the chromatic zeros
(those near to the origin) have support for $Re(q) < 0$, but the locus 
${\cal B}$ itself only has support for $Re(q) \ge 0$.  
We have encountered this type of situation in 
earlier work \cite{pg,bcc}.  The property that ${\cal B}$ only has support for
$Re(q) \ge 0$ can be demonstrated
by carrying out a Taylor series expansion of the degeneracy equation 
$|\lambda_{st4,10}|=|\lambda_{st4,25}|$ near the origin, which is, 
numerically, 
\beq
|44.1-52.0q+27.6q^2+O(q^3)|=|44.1-32.2q+8.8q^2+O(q^3)| \ . 
\label{lam10lam25degeneq}
\eeq
More generally, consider a degeneracy equation determining a curve on ${\cal
B}$ which, in the vicinity of the origin $q=0$, has the form 
\beq
|a_0+a_1q+a_2q^2+O(q^3)|=|a_0+b_1q+b_2q^2+O(q^3)|
\label{degeneqorigin}
\eeq
where the coefficients $a_i$ and $b_i$ are real and nonzero, 
$a_1 \ne b_1$, and, without loss of generality, we can take $a_0 > 0$. 
Writing $q$ in polar coordinates as $q=re^{i\theta}$ and expanding for small 
$r$, eq. (\ref{degeneqorigin}) reduces, to order $r$, to the equation 
$a_0(a_1-b_1)r\cos \theta=0$, which has as its solution $\theta=\pm \pi/2$.
Thus the curve ${\cal B}$ defined by a degeneracy equation of the form 
(\ref{degeneqorigin}) passes through the origin vertically.  In order to
determine in which direction (right or left) 
the curve bends away from the vertical as one moves away
from the origin, let us write $q=q_{_R}+iq_{_I}$ where 
$q_{_R}$ and $q_{_I}$ are real, with $q_{_R}^2+q_{_I}^2 = r^2 \ll 1$
Substituting, expanding, and using the fact that the curve ${\cal B}$ passes
vertically through the origin so that near this point $|q_R|$ is small compared
with $|q_I|$, we find, to this order, 
\beq
q_{_R} = \frac{[2a_0(a_2-b_2)+b_1^2-a_1^2]q_{_I}^2}{2a_0(a_1-b_1)} \ .
\label{qr}
\eeq
Thus if the right-hand side of this equation is positive (negative), the curve
${\cal B}$ bends to the right (left) into the half-plane with $Re(q) > 0$ 
($Re(q) < 0$) as one moves away from the origin.  
For the degeneracy equation (\ref{lam10lam25degeneq}), the right-hand side of 
eq. (\ref{qr}) is positive, so ${\cal B}$ bends to the right near the origin. 
As is evident from Fig. \ref{sqpxpy4}, as one moves farther away from the
origin, the curve ${\cal B}$ 
bends farther to the right, so that ${\cal B}$ has no support for $Re(q) < 0$. 
This is to be contrasted with the situation for (the $L_x \to \infty$ limit of)
sufficiently wide strips with cyclic or M\"obius boundary conditions (the $L_x
\to \infty$ limits of a given strip with cyclic boundary conditions is the same
as the limit with M\"obius boundary conditions), 
where it was found that for widths 
$L_y=3,4$ for the square lattice \cite{wcy,s4} and for width $L_y=4$ for
the triangular lattice \cite{t}, 
${\cal B}$ did have some support for $Re(q) < 0$.
A comparison of some properties of ${\cal B}$ in the present case and for other
strips with periodic or orientation-reversing periodic longitudinal boundary 
conditions is given in Table \ref{proptable}. 

\begin{table}
\caption{\footnotesize{Properties of $P$, $W$, and ${\cal B}$ for strip graphs
$G_s$ of the square (sq) and triangular (tri) lattices with periodic
longitudinal boundary conditions $(FBC_y,(T)PBC_x)$
(cyclic and M\"obius) and $(PBC_y,(T)PBC_x)$ (torus and Klein bottle). 
The properties apply
for a given strip of type $G_s$ of size $L_y \times L_x$; some apply for
arbitrary $L_x$, such as $N_{G_s,\lambda}$, while others apply for the
infinite-length limit, such as the properties of the locus ${\cal B}$.
The entry 37(38) for $N_{G_s,\lambda}$ means that 
$P$ has 37 different $\lambda_j$'s,
but the coloring matrix also has a zero eigenvalue that does not contribute to
$P$. The column denoted eqs. describes
the numbers and degrees of the algebraic equations giving the
$\lambda_{G_s,j}$; for example, $\{9(1),6(2),4(3)\}$ indicates that there are
nine linear equations, six quadratic equations and four cubic equation.  The 
column denoted BCR lists the points at which ${\cal B}$ crosses the real $q$ 
axis; here the largest of these is $q_c$ for the given family $G_s$. 
Column labelled ``SN'' refers to whether ${\cal B}$ has
\underline{s}upport for \underline{n}egative $Re(q)$, indicated as yes (y) or
no (n).}}
\begin{center}
\begin{tabular}{|c|c|c|c|c|c|c|c|c|c|}
\hline\hline $G_s$ & $L_y$ & $BC_y$ & $BC_x$ & $N_{G_s,\lambda}$ & eqs. & BCR \
& $q_c$ & SN & ref. \\ \hline\hline
sq  & 3 & P  & P   & 8 & \{8(1)\}             & 0, \ 2, \ 3 & 3 & n  & 16 \\
\hline
sq  & 3 & P  & TP  & 5 & \{5(1)\}             & 0, \ 2, \ 3 & 3 & n  & 16 \\
\hline
sq  & 4 & P  & P   & 33 & \{9(1),6(2),4(3)\}   & 0, \ 2, \ 2.78 & 2.78 & n & 
here \\ \hline
sq  & 4 & P  & TP  & 22 &  \{7(1),3(2),3(3)\}  & 0, \ 2, \ 2.78 & 2.78 & n &
 here \\ \hline\hline
sq  & 1 & F  &  P  & 2  & \{2(1)\}             & 0, \ 2  & 2 & n & 8 \\ \hline
sq  & 2 & F  & (T)P& 4  & \{4(1)\}             & 0, \ 2  & 2 & n & 8  \\ \hline
sq  & 3 & F  & (T)P& 10 & \{5(1),1(2),1(3)\}   & 0, \ 2, \ 2.34 & 2.34 & y & 
19-21 \\ \hline
sq  & 4 & F  & (T)P& 26 & \{4(1),1(2),2(3),1(4),2(5)\} & 0, \ 2, \ 2.49 & 
2.49 & y & 25 \\ \hline\hline
tri & 3 & P  & P  & 11 & \{5(1),3(2)\}        & 0, \ 2, \ 3.72 & 3.72 & n  &  
17 \\ \hline
tri & 3 & P  & TP & 5  & \{5(1)\}             & 0, \ 2, \ 3.72 & 3.72 & n  & 
17 \\ \hline
tri & 4 & P  & P  & 37(38) & \{5(1),4(2),2(3),3(4),1(6)\} & 0, \ 2, 4 & 4 & 
n & here \\ \hline
tri & 4 & P  & TP & 12(13) & \{4(1),1(2),2(3)\} & 0, \ 2, 4 & 4 & n & here \\ 
\hline\hline
tri & 2 & F  & (T)P & 4  & \{2(1),1(2)\} & 0, \ 2, \ 3 & 3 & n & 20 \\ \hline
tri & 3 & F  & (T)P & 10 & \{3(1),2(2),1(3)\} & 0, \ 2, \ 3 & 3 & n & 17 
\\ \hline
tri & 4 & F  &    P & 26 & \{1(1),2(4),1(8),1(9)\} & 0, \ 2, \ 3, \ 3.23 & 
3.23 & y & 17 \\
\hline\hline
\end{tabular}
\end{center}
\label{proptable}
\end{table}

\section{$L_y=4$ Strip of the Square Lattice with $(PBC_y,TPBC_x)$}

In general, for 
the strip graph of the square lattice with even width $L_y$ and 
$(PBC_y,TPBC_y)$, i.e., Klein bottle boundary conditions, we find that 
$\chi=4$ if $L_x=2$ and, for $L_x \ge 3$, 
\beq
\chi(sq, L_y \times L_x,PBC_y,TPBC_x)  = \cases{ 2 & if $L_x$ is odd \cr
                                              3 & if $L_x$ is even}
\label{chisqkb}
\eeq
For this strip (labelled $sk4$) we calculate that $N_{sk4,\lambda}=22$ and 
\beq
P(sq, 4 \times L_x,PBC_y,TPBC_x,q) = \sum_{j=1}^{22} c_{sk4,j} \ .
(\lambda_{sk4,j})^{L_x}
\label{psqkb}
\eeq
The nonzero terms $\lambda_{sk4,j}$ are identical to a subset of the terms 
$\lambda_{st4,j}$'s for the same strip with torus boundary conditions.
The 11 terms that occur in the chromatic polynomial (\ref{psqtorus}) for 
toroidal boundary conditions but are absent in the chromatic polynomial 
(\ref{psqkb}) for the Klein bottle case are 
\beq
\lambda_{st4,j} \ , j=4,5,12,13,16,17,20,21,31,32,33 \ . 
\label{lammissing}
\eeq
We have 
\beq
\lambda_{sk4,j}=\lambda_{st4,j} \quad {\rm for} \ \ 1 \le j \le 3
\label{lamsk41}
\eeq
\beq
\lambda_{sk4,j}=\lambda_{st4,j+2} \quad {\rm for} \ \ 4 \le j \le 9
\label{lamsk44}
\eeq
\beq
\lambda_{sk4,j}=\lambda_{st4,j+4}  \quad {\rm for} \ \ j=10,11
\label{lamsk410}
\eeq
\beq
\lambda_{sk4,j}=\lambda_{st4,j+6}  \quad {\rm for} \ \ j=12,13
\label{lamsk412}
\eeq
\beq
\lambda_{sk4,j}=\lambda_{st4,j+8}  \quad {\rm for} \ \ 14 \le j \le 22 \ . 
\label{lamsk414}
\eeq

The corresponding coefficients are 
\beq
c_{sk4,1}=1
\label{csk41}
\eeq
\beq
c_{sk4,2}=q-1
\label{csk42}
\eeq
\beq
c_{sk4,3}=\frac{1}{2}(q-1)(q-2)
\label{csk43}
\eeq
\beq
c_{sk4,4}=-(q-1)
\label{csk44}
\eeq
\beq
c_{sk4,5}=-c_{st4,7}=-\frac{1}{2}(q-1)(q-2)
\label{csk45}
\eeq
\beq
c_{sk4,6}=c_{st4,8}=\frac{1}{2}q(q-3)
\label{csk46}
\eeq
\beq
c_{sk4,7}=-c_{st4,9}=-1
\label{csk47}
\eeq
\beq
c_{sk4,j}=c_{st4,j+2}=1 \quad {\rm for} \ \ j=8,9
\label{csk489}
\eeq
\beq
c_{sk4,j}=-c_{st4,j+4}=-\frac{1}{2}q(q-3) \quad {\rm for} \ \ j=10,11
\label{csk41011}
\eeq
\beq
c_{sk4,j}=-c_{st4,j+6}=-\frac{1}{2}(q-1)(q-2) \quad {\rm for} \ \ j=12,13
\label{csk41213}
\eeq
\beq
c_{sk4,j}=-c_{st4,j+8}=-(q-1) \quad {\rm for} \ \ 14 \le j \le 16
\label{csk41416}
\eeq
\beq
c_{sk4,j}=c_{st4,j+8}=q-1 \quad {\rm for} \ \ 17 \le j \le 19
\label{csk41719}
\eeq
\beq
c_{sk4,j}=c_{st4,j+8}=\frac{1}{2}q(q-3) \quad {\rm for} \ \ 20 \le j \le 22 \ .
\label{csk42022}
\eeq
The sum of these coefficients is zero, as dictated by the $L_y=4$ special case
of the general result (\ref{csumkb}) above. 

Because none of the terms $\lambda_{st4,j}$ in (\ref{lammissing}) that is
present in (\ref{psqtorus}) and absent in (\ref{psqkb}) is dominant, it follows
that in the limit $L_x \to \infty$, the $W$ functions are the same for both of
these boundary conditions, and hence, so is the singular locus ${\cal B}$.
Below we shall prove in general that this must be the case; that is, in the
limit $L_x \to \infty$, a strip of the square (or triangular) lattice of width
$L_y$ with $(PBC_y,PBC_x)$ (torus) boundary conditions yields the same $W$
function and singular locus ${\cal B}$ as the corresponding strip with
$(PBC_y,TPBC_x)$ (Klein bottle) boundary conditions.

\section{$L_y=4$ Strip of the Triangular Lattice with $(PBC_y,PBC_x)$}

By similar methods, we have calculated the chromatic polynomials for strips 
of the triangular lattice with width $L_y=4$, arbitrarily great length $L_x$, 
and 
torus boundary conditions (labelled $tt4$).  In general, for a strip of the
triangular lattice of size $L_y \times L_x$ with toroidal boundary conditions,
for $L_y \ge 3$ and $L_x \ge 3$, the chromatic number is given by 
\beq
\chi(tri,L_y \times L_x,PBC_y,PBC_x) = \cases{ 3 & if $L_y=0$ \ mod 3 \ and 
$L_x=0$ \ mod 3 \cr
        4 & otherwise}
\label{chitritorus}
\eeq
Thus, in the present case, $\chi=4$, independent of $L_x$.  In the notation 
of eq. (\ref{pgsum}) we find $N_{tt4,\lambda}=37$ and 
\beq
P(tri, 4 \times L_x,PBC_y,PBC_x,q) = \sum_{j=1}^{37} c_{tt4,j} 
(\lambda_{tt4,j})^{L_x}
\label{ptorustri}
\eeq
where
\beq
\lambda_{tt4,1}=2
\label{lamtritor1}
\eeq
\beq
\lambda_{tt4,(2,3)}=\sqrt{2}e^{\pm i\pi/4} 
\label{lamtritor2}
\eeq
\beq
\lambda_{tt4,4}=2(3-q)
\label{lamtritor4}
\eeq
\beq
\lambda_{tt4,5}=3-q
\label{lamtritor5}
\eeq
\beq
\lambda_{tt4,6}=-2(2q-9)
\label{lamtritor6}
\eeq
\beq
\lambda_{tt4,7}=2(q-3)^2
\label{lamtritor7}
\eeq
\beqs
\lambda_{tt4,(8,9)} & = & \frac{(q-3)}{2}\biggl [ q^3-9q^2+33q-48 \cr\cr
& & \pm (q-4)(q^4-10q^3+43q^2-106q+129)^{1/2} \biggr ] 
\label{lamtritor89}
\eeqs
\beq
\lambda_{tt4,(10,11)}=\pm i \sqrt{3}\ (q-3)
\label{lamtritor1011}
\eeq
\beq
\lambda_{tt4,(12,13)}=\pm i (q-2) \sqrt{2(q-3)(q-4)} \ . 
\label{lamtritor1213}
\eeq
The $\lambda_{tt4,j}$'s for $14 \le j \le 19$ are roots of two cubic 
equations,
\beqs
& & \xi^3+2(q^3-12q^2+51q-75)\xi^2 \cr\cr
& & - 4(q-3)^3(q^2-7q+13)\xi-8(q-3)^4(q^2-5q+5)=0
\label{eqcub1tritor}
\eeqs
with roots $\lambda_{tt4,j}$, $j=14,15,16$, and 
\beqs
& & \xi^3 - 2(2q^2-17q+39)\xi^2 + 2(q^4-17q^3+100q^2-244q+214)\xi \cr\cr
& & + 4(q-3)(q^4-11q^3+44q^2-76q+46)=0 
\label{eqcub2tritor}
\eeqs
with roots $\lambda_{tt4,j}$, $j=17,18,19$. 
The $\lambda_{tt4,j}$'s for $20 \le j \le 31$ are roots of three quartic
equations,
\beqs
& & \xi^4 + 2(q^3-9q^2+29q-34)\xi^3 + 2(q^3-9q^2+29q-34)^2\xi^2 \cr\cr
& & + 4(q-3)^2(q^2-5q+5)(q^3-9q^2+29q-34)\xi + 4(q-3)^4(q^2-5q+5)^2=0 \cr\cr
& & 
\label{eqquartic1tritor}
\eeqs
with roots $\lambda_{tt4,j}$, $20 \le j \le 23$, 
\beqs
& & \xi^4 - 2(q^2-7q+14)\xi^3 + 2(q^2-7q+14)^2\xi^2 \cr\cr
& & +4(q-3)(q^2-7q+14)(q^2-6q+7)\xi+4(q-3)^2(q^2-6q+7)^2=0 \cr\cr
& & 
\label{eqquartic2tritor}
\eeqs
with roots $\lambda_{tt4,j}$, $24 \le j \le 27$, and 
\beqs
& & \xi^4 + 2(3q-11)\xi^3 + 2(3q-11)^2\xi^2 \cr\cr
& & +2(3q-11)(3q^2-18q+23)\xi+(3q^2-18q+23)^2=0 \cr\cr
& & 
\label{eqquartic3tritor}
\eeqs
with roots $\lambda_{tt4,j}$, $28 \le j \le 31$. 
Finally, the $\lambda_{tt4,j}$'s for $32 \le j \le 37$ are roots of an equation
of degree six: 
\beqs
& & \xi^6 - 2(q-5)(2q-7)\xi^5 + 2(q-5)^2(2q-7)^2\xi^4 \cr\cr
& & + 8(q-4)^2(3q^3-29q^2+89q-85)\xi^3 + 4(3q^3-28q^2+84q-79)^2\xi^2 \cr\cr
& & + 8(q-3)^2(q^2-5q+5)(3q^3-28q^2+84q-79)\xi + 8(q-3)^4(q^2-5q+5)^2=0 \ . 
\cr\cr
& & 
\label{eqsixtritor}
\eeqs
Each of the three quartic equations above has roots of the form 
$a_\ell e^{\pm i\pi/4}$, $b_\ell e^{\pm i\pi/4}$, where $\ell=1,2,3$ indexes
the quartic equation, so 
\beq
\lambda_{tt4,j}=a_1 e^{\pm i\pi/4} \quad {\rm for} \ \ j=20,21
\label{lamtritor2021}
\eeq
\beq
\lambda_{tt4,j}=b_1 e^{\pm i\pi/4} \quad {\rm for} \ \ j=22,23
\label{lamtritor2223}
\eeq
\beq
\lambda_{tt4,j}=a_2 e^{\pm i\pi/4} \quad {\rm for} \ \ j=24,25
\label{lamtritor2425}
\eeq
\beq
\lambda_{tt4,j}=b_2 e^{\pm i\pi/4} \quad {\rm for} \ \ j=26,27
\label{lamtritor2627}
\eeq
\beq
\lambda_{tt4,j}=a_3 e^{\pm i\pi/4} \quad {\rm for} \ \ j=28,29
\label{lamtritor2829}
\eeq
\beq
\lambda_{tt4,j}=b_3 e^{\pm i\pi/4} \quad {\rm for} \ \ j=30,31
\label{lamtritor3031}
\eeq
where the values of $a_\ell$ and $b_\ell$, $\ell=1,2,3$ are determined by these
quartic equations. Similarly, the roots of the 
sixth-order equation are of the form $c_\ell e^{\pm i\pi/4}$, 
$\ell=1,2,3$, i.e., 
\beq
\lambda_{tt4,j}=c_1 e^{\pm i\pi/4} \quad {\rm for} \ \ j=32,33
\label{lamtritor3233}
\eeq
\beq
\lambda_{tt4,j}=c_2 e^{\pm i\pi/4} \quad {\rm for} \ \ j=34,35
\label{lamtritor3435}
\eeq
\beq
\lambda_{tt4,j}=c_3 e^{\pm i\pi/4} \quad {\rm for} \ \ j=36,37
\label{lamtritor3637}
\eeq
where the values of $c_\ell$, $\ell=1,2,3$ follow from eq. 
(\ref{eqsixtritor}). 
Below we shall comment further on these phase factors. 

The corresponding coefficients are
\beq
c_{tt4,1}=\frac{1}{4}q(q-2)(q-3)^2
\label{ctt4_1}
\eeq
\beq
c_{tt4,2}=c_{tt4,3}=\frac{1}{12}(q-1)(q-2)(3q^2-11q-6)
\label{cct4_23}
\eeq
\beq
c_{tt4,j}=\frac{1}{2}(q-1)(q-2) \quad {\rm for} \ \ j=4,7 \ \ {\rm and} \ \ 
32 \le j \le 37
\label{ctt4_4}
\eeq
\beq
c_{tt4,5}=\frac{2}{3}q(q-2)(q-4)
\label{ctt4_5}
\eeq
\beq
c_{tt4,j}=\frac{1}{3}q(q-2)(q-4) \quad {\rm for} \ \ j=10,11 \ \ {\rm and} \ \ 
28 \le j \le 31
\label{ctritor2831}
\eeq
\beq
c_{tt4,6}=\frac{1}{3}(q-1)(q^2-5q+3)
\label{ctt4_6}
\eeq
\beq
c_{tt4,8}=c_{tt4,9}=1
\label{ctritor8}
\eeq
\beq
c_{tt4,j}=\frac{1}{2}q(q-3)  \quad {\rm for} \ \ j=12,13,17,18,19 \ \ 
{\rm and} \ \ 24 \le j \le 27
\label{ctt4_1213}
\eeq
\beq
c_{tt4,j} = q-1 \quad {\rm for} \ \ j=14,15,16 \ \ {\rm and} \ \ 
20 \le j \le 23 \ .
\label{ctritor1416}
\eeq
Formally, we have also found a zero eigenvalue, 
\beq
\lambda_{tt4,38}=0
\label{lamtritor38}
\eeq
with coefficient (multiplicity)
\beq
c_{tt4,38}=\frac{1}{12}q(q-1)(3q^2-17q+40) \ .
\label{ctritor38}
\eeq
Although this term does not contribute to the chromatic polynomial 
(\ref{pgsum}), the corresponding coefficient does contribute to the sum of
multiplicities, i.e. to the total dimension of the space of coloring
configurations, given by (\ref{csumtor}).  The sum of all of the coefficients,
including that corresponding to the zero eigenvalue, is equal to
$P(C_4,q)=q(q-1)(q^2-3q+3)$, which is an $L_y=4$ special case of
(\ref{csumtor}). 

\begin{figure}[hbtp]
\centering
\leavevmode
\epsfxsize=3.0in
\begin{center}
\leavevmode
\epsffile{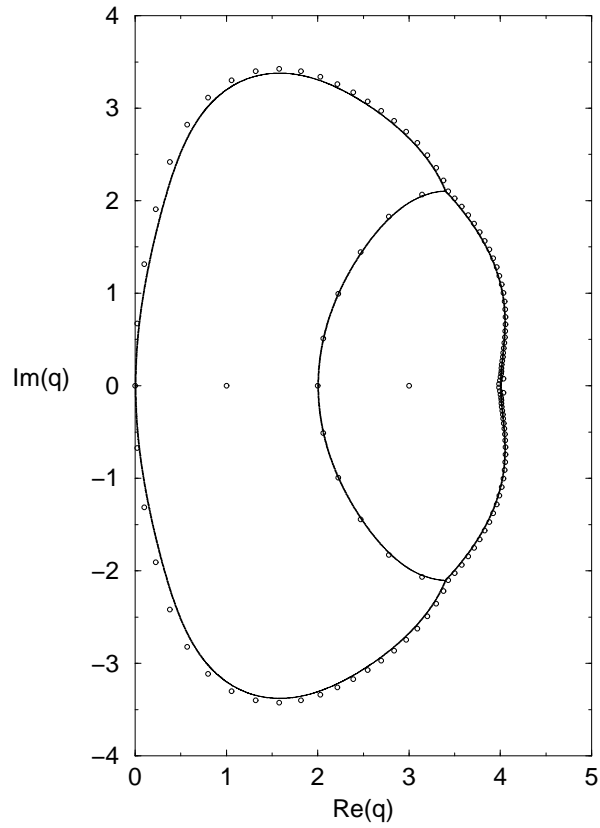}
\end{center}
\caption{\footnotesize{Singular locus ${\cal B}$ for the $L_x \to \infty$ limit
of the strip of the triangular lattice with $L_y=4$ and toroidal boundary
conditions. For comparison, chromatic zeros are shown for $L_x=30$ (i.e.,
$n=120$).}}
\label{tpxpy4}
\end{figure}

The singular locus ${\cal B}$ for the $L_x \to \infty$ limit of the strip of
the triangular lattice with $L_y=4$ and toroidal boundary conditions is shown
in Fig. \ref{tpxpy4}.  For comparison, chromatic zeros are calculated and
shown for length $L_x=30$ (i.e., $n=120$ vertices). 
The locus ${\cal B}$ crosses the real axis at the points $q=0$, $q=2$, and at
the maximal point $q=q_c$, where
\beq
q_c=4  \quad {\rm for} \ \ \{G\}=(tri,4 \times L_x,PBC_y,PBC_x) .
\label{qctri}
\eeq
At this point there are several degeneracies of magnitudes of eigenvalues;
these occur for $\lambda_j$ with $j=1,4,6,7,8,9$ and $14 \le j \le 19$. 

As is evident from Fig. \ref{tpxpy4}, the locus ${\cal B}$ separates the $q$
plane into different regions including the following (we use the same symbols
as for the $L_y=4$ toroidal strip of the square lattice, but it is understood
that the regions are specific to this section): (i) $R_1$, containing the
semi-infinite intervals $q > 4$ and $q < 0$ on the real axis and extending
outward to infinite $|q|$, (ii) $R_2$ containing the interval $2 < q < 4$, and
(iii) $R_3$ containing the real interval $0 < q < 2$ Again, the (nonzero)
density of chromatic zeros has the smallest values on the curve separating
regions $R_1$ and $R_3$ in the vicinity of the point $q=0$ and on the curve
separating regions $R_2$ and $R_3$ in the vicinity of the point $q=2$.

In region $R_1$, $\lambda_{tt4,8}$ is the dominant $\lambda_{G,j}$, so
\beq
W = (\lambda_{tt4,8})^{1/4} \ , \quad q \in R_1 \ .
\label{wtrir1}
\eeq
This is the same as $W$ for the corresponding $L_x \to \infty$ limit of the
strip of the triangular lattice with the same width $L_y=4$ and cylindrical
$(PBC_y,FBC_x)$ boundary conditions, calculated in \cite{strip2}.  

In region $R_2$,
\beq
|W| = |\lambda_{tt4,17}|^{1/4} \ , \quad q \in R_2
\label{wtrir2}
\eeq
where $\lambda_{tt3,17}$ is the root of the cubic equation 
(\ref{eqcub2tritor}) that
has the maximal magnitude for $2 < q < 4$.  In region $R_3$,
\beq
|W|=|\lambda_{tt4,14}|^{1/4} \ , \quad q \in R_3 \ .
\label{wtrir3}
\eeq
There are no other regions containing nonzero intervals of the real axis
besides $R_j$, $j=1,2,3$.  However, our previous calculations for various
families of graphs \cite{wcy,t} have shown that ${\cal B}$ can include pairs of
extremely small complex-conjugate sliver regions. We have not made an
exhaustive search for these in the present case.  

Corresponding to eq. (\ref{wsqtor4qgeqc}) for the toroidal or Klein bottle and
cylindrical strips of the square lattice, we have 
\beq
W(tri,4 \times \infty,PBC_y,(T)PBC_x,q)=
W(tri,4 \times \infty,PBC_y,FBC_x,q) \quad 
{\rm for} \ \ q \ge 4 \ . 
\label{wtritor4qgeqc}
\eeq
Hence the values of $W(tri,4 \times \infty,PBC_y,FBC_x,q)$ for various values
of $q \ge 4$ given in \cite{s4} (see also \cite{t}) are also applicable here. 
For low integral values of $q$ we list the values of $|W(q)|$ for this strip in
Table \ref{wtriinternal}, 
together with corresponding values given in \cite{s4,t} 
for $W$ in the $L_x \to \infty$ limit of the $L_y=3$ strip with
$(PBC_y,(T)PBC_x)$. 

\begin{table}
\caption{\footnotesize{Values of $W(tri,L_y \times \infty,PBC_y,(T)PBC_x,q)$
for low integral $q$ and for respective $q_c$.}}
\begin{center}
\begin{tabular}{|c|c|c|c|c|c|c|c|c|}
\hline\hline
$L_y$ & $BC_y$ & $BC_x$ & $|W_{q=0}|$ & $|W_{q=1}|$ & $|W_{q=2}|$ &
$|W_{q=3}|$ & $q_c$ & $W_{q=q_c}$ \\
\hline\hline
3  & P & (T)P & 3.17   & 2.62    & 2      & 1.71   & 3.72  & 1.41  \\ \hline
4  & P & (T)P & 3.44   & 2.86    & 2.25   & 1.83   & 4     & 1.19  \\ 
\hline\hline
\end{tabular}
\end{center}
\label{wtriinternal}
\end{table}

The locus 
${\cal B}$ only has support for $Re(q) \ge 0$.  This can be demonstrated 
by carrying out a Taylor series expansion of the degeneracy equation
$|\lambda_{tt4,8}|=|\lambda_{tt4,14}|$ near the origin, which is, numerically,
\beq 
|140.15-141.25q+57.6q^2+O(q^3)|=|140.15-93.4q+21.9q^2+O(q^3)| \ .
\label{lam8lam14degeneq}
\eeq
This equation is of the form (\ref{degeneqorigin}), and, using eq. (\ref{qr}),
we verify that ${\cal B}$ bends to the right as one moves away from the 
origin.  Farther away from the origin, one can see from Fig. \ref{tpxpy4} that
${\cal B}$ continues to move into the half-plane with $Re(q) > 0$, so that the
conclusion stated above follows, that this locus has no support for $Re(q) <
0$. 

\section{$L_y=4$ Strip of the Triangular Lattice with $(PBC_y,TPBC_x)$}

The strip of the triangular lattice with width $L_y=4$, arbitrarily great 
length $L_x$, and $(PBC_y,TPBC_x)=$ Klein bottle boundary conditions, 
labelled $tk4$, has (for $L_x \ge 2$) chromatic number
\beq
\chi(tri,4 \times L_x,PBC_y,TPBC_x) = \cases{4  & if $L_x$ is even \cr
        5 & if $L_x$ is odd}
\label{chitrikb}
\eeq
In the notation of eq. (\ref{pgsum}) we find $N_{tk4,\lambda}=12$ and
\beq
P(tri, 4 \times L_x,PBC_y,TPBC_x,q) = \sum_{j=1}^{12} c_{tk4,j} 
(\lambda_{tk4,j})^{L_x}
\label{ptk4}
\eeq
where
\beq
\lambda_{tk4,1}=\lambda_{tt4,1}=2 
\label{lamtk41}
\eeq
\beq
\lambda_{tk4,2}=\lambda_{tt4,4}=2(3-q)
\label{lamtk42}
\eeq
\beq
\lambda_{tk4,3}=\lambda_{tt4,6}=-2(2q-9)
\label{lamtk43}
\eeq
\beq
\lambda_{tk4,4}=\lambda_{tt4,7}=2(q-3)^2
\label{lamtk44}
\eeq
\beq
\lambda_{tk4,j}=\lambda_{tt4,j+3} \quad {\rm for} \ \ j=5,6
\label{lamtk56}
\eeq
\beq
\lambda_{tk4,j}=\lambda_{tt4,j+7} \quad {\rm for} \ \ 7 \le j \le 12 \ . 
\label{lamtk4712}
\eeq

The corresponding coefficients are 
\beq
c_{tk4,1}=\frac{1}{2}q(q-2)(q-3)
\label{ctk41}
\eeq
\beq
c_{tk4,2}=c_{tt4,4}=\frac{1}{2}(q-1)(q-2)
\label{ctk42}
\eeq
\beq
c_{tk4,3}=-(q-1)
\label{ctk43}
\eeq
\beq
c_{tk4,4}=-c_{tt4,7}=-\frac{1}{2}(q-1)(q-2)
\label{ctk44}
\eeq
\beq
c_{tk4,5}=c_{tk4,6}=c_{tt4,8}=c_{tt4,9}=1
\label{ctk456}
\eeq
\beq
c_{tk4,j}=c_{tt4,j+7}=q-1 \quad {\rm for} \ \ j=7,8,9
\label{ctk479}
\eeq
\beq
c_{tk4,j}=c_{tt4,j+7}=\frac{1}{2}q(q-3) \quad {\rm for} \ \ j=10,11,12 \ . 
\label{ctk41012}
\eeq
The coloring matrix also has another eigenvalue, namely, 
\beq
\lambda_{tk4,13}=0
\label{lamtk413}
\eeq
with multiplicity
\beq
c_{tk4,13}=-\frac{1}{2}q(q-1)^2 \ .
\label{ctk413}
\eeq
Hence, the total number of distinct eigenvalues of the coloring matrix for this
strip is $N_{tk4,\lambda,tot}=N_{tk4,\lambda}+1=13$. 
The sum of all of the coefficients, including that for the zero eigenvalue, is
zero; this is an $L_y=4$ special case of (\ref{csumkb}). 

\section{Cyclic and Toroidal Crossing-Subgraph Strips of the Square Lattice}

\subsection{General} 

It is worthwhile to include here some results on certain related families of
strip graphs since these give insight into the structure of the chromatic
polynomials for the various strips with longitudinal boundary conditions which
are periodic or periodic with reversed orientation. Let us consider first 
a strip of the square lattice of fixed width $L_y$ and 
arbitrarily great length $L_x$ constructed as follows.  As before, 
the longitudinal (horizontal) direction on the strip to be $x$ and the
transverse (vertical) direction to be $y$.  Label the vertices of
two successive transverse slices of the strip, 
starting at the top as $(1,2,..,L_y)$ and
$(1^\prime,2^\prime,...,L_y^\prime)$.  First, consider the case of free
transverse boundary conditions, for which these transverse slices of the strip
are line (path) graphs with $L_y$ vertices. Connect these with edges linking
vertices 1 to $L_y^\prime$, 2 to $(L_y-1)^\prime$, 3 to $(L_y-2)^\prime$, 
and so forth.  For example,
for $L_y=2$, we connect 1 to $2^\prime$ and 2 to $1^\prime$; for $L_y=3$, we
connect 1 to $3^\prime$, 2 to $2^\prime$, and 3 to $1^\prime$, etc. for other
values of $L_y$.   An example of this crossing-subgraph strip of the square
lattice of width $L_y=3$ is given in Fig. \ref{cgraphsq}(a). 

\vspace*{5mm}
\unitlength 1.3mm
\begin{center}
\begin{picture}(40,20)
\multiput(0,0)(10,0){5}{\circle*{2}}
\multiput(0,10)(10,0){5}{\circle*{2}}
\multiput(0,20)(10,0){5}{\circle*{2}}
\multiput(0,0)(10,0){5}{\line(0,1){20}}
\put(0,10){\line(1,0){40}}
\multiput(0,0)(10,0){4}{\line(1,2){10}}
\multiput(0,20)(10,0){4}{\line(1,-2){10}}
\put(-2,-2){\makebox(0,0){3}}
\put(8,-2){\makebox(0,0){6}}
\put(18,-2){\makebox(0,0){9}}
\put(28,-2){\makebox(0,0){12}}
\put(38,-2){\makebox(0,0){3}}
\put(-2,12){\makebox(0,0){2}}
\put(8,12){\makebox(0,0){5}}
\put(18,12){\makebox(0,0){8}}
\put(28,12){\makebox(0,0){11}}
\put(38,12){\makebox(0,0){2}}
\put(-2,22){\makebox(0,0){1}}
\put(8,22){\makebox(0,0){4}}
\put(18,22){\makebox(0,0){7}}
\put(28,22){\makebox(0,0){10}}
\put(38,22){\makebox(0,0){1}}
\put(20,-8){\makebox(0,0){(a)}}
\end{picture}
\end{center}

\vspace*{20mm}

\begin{center}
\begin{picture}(40,30)
\multiput(0,0)(10,0){5}{\circle*{2}}
\multiput(0,10)(10,0){5}{\circle*{2}}
\multiput(0,20)(10,0){5}{\circle*{2}}
\multiput(0,30)(10,0){5}{\circle*{2}}
\multiput(0,0)(10,0){5}{\line(0,1){30}}
\put(0,20){\line(1,0){40}}
\multiput(0,0)(10,0){4}{\line(1,1){10}}
\multiput(0,10)(10,0){4}{\line(1,-1){10}}
\put(-2,-2){\makebox(0,0){1}}
\put(8,-2){\makebox(0,0){4}}
\put(18,-2){\makebox(0,0){7}}
\put(28,-2){\makebox(0,0){10}}
\put(38,-2){\makebox(0,0){1}}
\put(-2,12){\makebox(0,0){3}}
\put(8,12){\makebox(0,0){6}}
\put(18,12){\makebox(0,0){9}}
\put(28,12){\makebox(0,0){12}}
\put(38,12){\makebox(0,0){3}}
\put(-2,22){\makebox(0,0){2}}
\put(8,22){\makebox(0,0){5}}
\put(18,22){\makebox(0,0){8}}
\put(28,22){\makebox(0,0){11}}
\put(38,22){\makebox(0,0){2}}
\put(-2,32){\makebox(0,0){1}}
\put(8,32){\makebox(0,0){4}}
\put(18,32){\makebox(0,0){7}}
\put(28,32){\makebox(0,0){10}}
\put(38,32){\makebox(0,0){1}}
\put(20,-8){\makebox(0,0){(b)}}
\end{picture}
\end{center}
\vspace*{5mm}
\begin{figure}[h]
\caption{\footnotesize{Illustrative crossing-subgraph strip graphs of the
square lattice with (a) $(FBC_y,PBC_x)=$ cyclic and (b) $(PBC_y,PBC_x)=$ 
toroidal type.  For these, $L_y=3$ and $L_x=4$. Vertices are indicated with 
$\bullet$ (points where edges cross without a symbol $\bullet$ are not
vertices.) }}
\label{cgraphsq}
\end{figure}

\noindent 

We impose periodic longitudinal boundary conditions.  
We shall denote this crossing-subgraph strip (labelled $cg$) of the square
($sq$) lattice as $cg(sq,L_y \times L_x, FBC_y,PBC_x)$.  We  observe that 
\beq
cg(sq,L_y \times L_x,FBC_y,PBC_x)= \cases{(sq,L_y \times L_x,FBC_y,PBC_x) & 
                                                  if $L_x$ is even \cr
                                      (sq,L_y \times L_x,FBC_y,TPBC_x) & 
                                                  if $L_x$ is odd} 
\label{cgidentity}
\eeq
That is, for even (odd) $L_x$, this crossing-subgraph strip 
reduces to the cyclic
(M\"obius) strip of the square lattice.  Secondly, consider the case where we
impose periodic transverse boundary conditions; we denote this toroidal
crossing-subgraph strip of the square lattice as 
$cg(sq,L_y \times L_x, PBC_y,PBC_x)$.  In this case the transverse slices are
circuit graphs with $L_y$ vertices.   An example of this toroidal
crossing-subgraph strip of the square lattice with width $L_y=3$ is shown in
Fig. \ref{cgraphsq}(b).  We have 
\beq
cg(sq,L_y \times L_x,PBC_y,PBC_x)= \cases{(sq,L_y \times L_x,PBC_y,PBC_x) &
                                                  if $L_x$ is even \cr
                                      (sq,L_y \times L_x,PBC_y,TPBC_x) &
                                                  if $L_x$ is odd}
\label{cgtkidentity}
\eeq
That is, for even (odd) $L_x$, this crossing-subgraph strip reduces to the 
strip of the square lattice with torus (Klein bottle) boundary conditions. 
Given the relations (\ref{cgidentity}) and (\ref{cgtkidentity}), 
it follows that a knowledge of the chromatic polynomial for the cyclic 
crossing-subgraph strip of width $L_y$ of
the square lattice is equivalent to a knowledge of the chromatic polynomials
for the strip of this lattice with width $L_y$ with both cyclic and M\"obius
boundary conditions and, similarly, a knowledge of the chromatic polynomial for
the toroidal crossing-subgraph strip of the square lattice is equivalent to a 
knowledge of the chromatic polynomials for the strip of this lattice with 
both torus and Klein bottle boundary conditions.  

The sum of the coefficients for the $L_y \times L_x$ cyclic crossing-subgraph
strip of the square lattice is
\beq
C_{cg(sq,L_y \times L_x,FBC_y,PBC_x)}=P(T_{L_y},q)
\label{csumcgsq}
\eeq
where $T_n$ is the tree graph on $n$ vertices, and $P(T_n,q)=q(q-1)^{n-1}$. 
The sum of the coefficients for the $L_y \times L_x$ toroidal crossing-subgraph
strip of the square or triangular lattice is
\beq
C_{cg(G_s,L_y \times L_x,PBC_y,PBC_x)}=P(C_{L_y},q) \quad {\rm for} \ \ 
G_s=sq,tri
\label{csumcgsqtri}
\eeq
as in (\ref{csumtor}).

We have carried out explicit
calculations of chromatic polynomials for a number of crossing-subgraph strips
(labelled $cg$) and have related the results to those for strips with
cyclic/M\"obius and torus/Klein bottle boundary conditions.  We concentrate
here on strips of the square lattice and discuss those of the triangular
lattice below. We find that a certain subset of the 
terms in (\ref{pgsum}) for the crossing-subgraph strips occur in
opposite-sign pairs of the form $\pm \lambda_{cg,j}$.  Consider the
coefficients for the $\pm \lambda_{cg,j}$'s in each pair: in some cases, these
are different, while in others they are the same.  Let us denote the number of
$\lambda_{cg,j}$'s comprising opposite-sign pairs such that the members of each
pair have different (the same) coefficients as $N_{cg,opd,\lambda}$
($N_{cg,ops,\lambda}$), respectively.  The number of remaining
$\lambda_{cg,j}$'s that are not members of an opposite-sign pair is denoted
$N_{cg,up,\lambda}$, where $up$ means ``unpaired''.  Clearly \beq
N_{cg,\lambda}=N_{cg,up,\lambda}+N_{cg,opd,\lambda}+N_{cg,ops,\lambda} \ .
\label{ncgeq}
\eeq
For the cyclic
and M\"obius strips that we have studied, we find $N_{cg,ops,\lambda}=0$, while
for torus and Klein bottle strips, $N_{cg,ops,\lambda}$ is, in general,
nonzero.  For even $L_x$, where, according to the identities 
(\ref{cgidentity}) and (\ref{cgtkidentity}), the 
cyclic (toroidal) crossing-subgraph strip reduces to the cyclic (toroidal) 
strip of the square lattice, two such opposite-sign terms reduce to a single
term as follows (the subscripts $jp,jm$ denote $j,\pm$) 
\beqs
c_{cg,cyc,jp}(\lambda_{cg,cyc,j})^{L_x} +
c_{cg,cyc,jm}(-\lambda_{cg,cyc,j})^{L_x} & = & 
(c_{cg,cyc,jp}+c_{cg,cyc,jm})(\lambda_{cg,cyc,j})^{L_x} \cr\cr
& = & c_{sq,cyc,j}(\lambda_{cg,cyc,j})^{L_x} \ .
\label{cgcoeffsum}
\eeqs
For odd $L_x=m$ the cyclic (toroidal) crossing-subgraph strip reduces to 
the M\"obius (Klein bottle) strip, and the pair of opposite-sign terms reduces
to a single term as follows: 
\beqs
c_{cg,cyc,jp}(\lambda_{cg,cyc,j})^{L_x} +
c_{cg,cyc,jm}(-\lambda_{cg,cyc,j})^{L_x} & = & 
(c_{cg,cyc,jp}-c_{cg,cyc,jm})(\lambda_{cg,cyc,j})^{L_x} \cr\cr
& = & c_{sq,Mb,j}(\lambda_{cg,cyc,j})^{L_x} 
\label{cgcoeffdiff}
\eeqs
where the subscript $Mb$ denotes M\"obius. 
In particular, if $\lambda_{cg,j}$ is one of the $N_{cg,ops,\lambda}$ terms
with $c_{cg,jp}=c_{cg,jm}$, then the terms in (\ref{cgcoeffdiff}) cancel each
other, leaving no contribution.  As noted above, in our studies, we have 
found that this can happen for 
Klein bottle strips, since $N_{cg,ops,\lambda} \ne
0$ for these, but not for M\"obius strips, since $N_{cg,ops,\lambda}=0$ for
these.  The inverse relations connecting the coefficients for the terms 
$\pm \lambda_{cg,cyc,j}$ in the chromatic polynomial for the crossing-subgraph 
cyclic strip to the coefficients $c_{sq,cyc,j}$ and $c_{sq,Mb,j}$ in the 
cyclic and M\"obius strips are thus 
\beq
c_{cg,cyc,jp} = \frac{1}{2}(c_{sq,cyc,j}+c_{sq,Mb,j})
\label{ccgplus}
\eeq
\beq
c_{cg,cyc,jm} = \frac{1}{2}(c_{sq,cyc,j}-c_{sq,Mb,j})
\label{ccgminus}
\eeq
and similarly, for the coefficients for the terms $\pm \lambda_{cg,torus,j}$ 
in the chromatic polynomial for the toroidal crossing-subgraph
strip in terms of the coefficients $c_{sq,torus,j}$ and $c_{sq,Kb,j}$ in the
torus and Klein bottle ($Kb$) strips, 
\beq
c_{cg,torus,jp} = \frac{1}{2}(c_{sq,torus,j}+c_{sq,Kb,j})
\label{ccgplustor}
\eeq
\beq
c_{cg,torus,jm} = \frac{1}{2}(c_{sq,torus,j}-c_{sq,Kb,j}) \ .
\label{ccgminuskb}
\eeq
 From these considerations, we derive the following general formula: 
\beq
N_{sq,L_y,cyc,\lambda}=N_{cg,sq,L_y,cyc,\lambda}-
\frac{1}{2}N_{cg,sq,L_y,cyc,opd,\lambda}
\label{nlamcgcyc}
\eeq
and, since $N_{cg,ops,\lambda}=0$ for the cyclic crossing-graph strips of the
square lattice that we have studied,
the same formula applies to the corresponding M\"obius strips with 
$N_{sq,L_y,cyc,\lambda}$ replaced by $N_{sq,L_y,Mb,\lambda}$.  Further, 
\beq
N_{sq,L_y,torus,\lambda}=N_{cg,sq,L_y,torus,\lambda}
-\frac{1}{2}N_{cg,sq,L_y,torus,opd,\lambda}
-\frac{1}{2}N_{cg,sq,L_y,torus,ops,\lambda}
\label{nlamcgtorus}
\eeq
\beq
N_{sq,L_y,Kb,\lambda}=N_{cg,sq,L_y,torus,\lambda}-
\frac{1}{2}N_{cg,sq,L_y,torus,opd,\lambda}
          -N_{cg,sq,L_y,torus,ops,\lambda} \ . 
\label{nlamcgkb}
\eeq
Thus, 
\beq
N_{sq,L_y,torus,\lambda}-N_{sq,L_y,Kb,\lambda}=
\frac{1}{2}N_{cg,sq,L_y,torus,ops,\lambda} \ .
\label{nlamcgtoruskb}
\eeq

In the limit $L_x \to \infty$, the cyclic or toroidal crossing-subgraph
strip of a given width $L_y$ yields a $W$ function via (\ref{w}) and hence a
singular locus ${\cal B}$.  An important theorem can be proved from this by
observing that we can take this limit using even or odd values of $L_x$;
in the even-$L_x$ case, we obtain the function $W$ and locus ${\cal
B}$ for the strip with torus boundary conditions, while in the odd-$L_x$ case,
we obtain the $W$ function and ${\cal B}$ for the strip with Klein bottle
boundary conditions.  Since the original limit exists, all three of these
limits must be the same.  This proves the following theorem:

\vspace{4mm}

\begin{flushleft} 

Theorem 1: \ The $W$ function and singular locus ${\cal B}$ are the same for 
the $L_x \to \infty$ limit of strip of the square lattice with width $L_y$ and
length $L_x$ whether one imposes $(PBC_y,PBC_x)$ or $(PBC_y,TPBC_x)$, i.e. 
torus or Klein bottle boundary conditions. 

\vspace{4mm}

Since the chromatic polynomials for these two sets of boundary conditions
involve different numbers of terms, this was not, {\it a priori} obvious.  This
feature was first noticed in \cite{tk} and was shown there to be a consequence
of the fact that none of the terms $\lambda_{st3,j}$ for the torus case
that
were absent in the Klein bottle case was dominant; here we have succeeded in
explaining why this had to be true; if it were not, then the respective loci
${\cal B}$ would be different, but this is impossible, as a consequence of our
present theorem.  Thus, a corollary to the theorem is

\vspace{4mm}

Corollary 1: \ Consider an $L_y \times L_x$ strip of the square lattice with
torus or Klein bottle boundary conditions, and denote the set of nonzero
eigenvalues that contribute to (\ref{pgsum}) for these two respective strips as
$\lambda_{stL_y,j}$, $j=1,..,N_{stL_y,\lambda}$ and $\lambda_{skL_y,j}$,
$j=1,..,N_{skL_y,\lambda}$. Focus on the set of eigenvalues $\lambda_{stL_y,j}$
that do not occur among the set $\lambda_{skL_y,j}$ (the number of these is
given by eq. (\ref{nlamcgtoruskb})); none of these can be dominant eigenvalues.

In a similar manner, one can prove that the locus ${\cal
B}$ for the $L_x \to \infty$ limits of the strips of the square lattice with
cyclic and M\"obius boundary conditions are the same without using as input the
identity of terms $\lambda_{sq,L_y,cyc,j}=\lambda_{sq,L_y,Mb,j}$ that we
have
observed in our studies \cite{pm,s4}.  Below we shall also prove a similar
theorem for strips of the triangular lattice with torus and Klein bottle
boundary conditions.

\end{flushleft} 

We recall that in \cite{cf} we observed from our work that the coefficients 
$c_{G,j}$ in (\ref{pgsum}) for cyclic and M\"obius strips of the square 
lattice are Chebyshev polynomials; in particular, for a given degree $d$
polynomial, there is a unique coefficient with this degree, and it is given by 
\beq
c^{(d)}=U_{2d}\Bigl (\frac{\sqrt{q}}{2} \Bigr )
\label{cd}
\eeq
where $U_n(x)$ is the Chebyshev polynomial of the second kind, defined
by
\beq
U_n(x) = \sum_{j=0}^{[\frac{n}{2}]} (-1)^j {n-j \choose j} (2x)^{n-2j}
\label{undef}
\eeq
where $[\frac{n}{2}]$ means the integral part of $\frac{n}{2}$. 
The first few of these coefficients are $c^{(0)}=1$, $c^{(1)}=q-1$, 
$c^{(2)}=q^2-3q+1$, $c^{(3)}=q^3-5q^2+6q-1$, 
etc.  We also found that the eigenvalues 
$\lambda_{G,j}$ were the same for cyclic and M\"obius strips of the square (and
triangular) lattices of a given width that we considered. 
We established the transformation rules
specifying how a coefficient of a given degree changes when one switches from
the cyclic to M\"obius strip of the square lattice \cite{cf}:
\beq
c^{(0)} \to \pm c^{(0)}
\label{cd0tran}
\eeq
\beq
c^{(2k)} \to \pm c^{(k-1)} \ ,  1 \le k \le
\Bigl [ \frac{L_y}{2} \Bigr ]
\label{cdeventran}
\eeq
\beq
c^{(2k+1)} \to \pm c^{(k+1)} \ ,  0 \le k \le
\Bigl [ \frac{L_y-1}{2} \Bigr ] \ . 
\label{cdoddtran}
\eeq
Following the notation of \cite{cf}, denote the number of terms (eigenvalues) 
$\lambda_{G,j}$ with coefficients $c^{(d)}$ as $n_P(L_y,d)$.  We concentrate on
the case of the square strip here and suppress the $sq$ in the notation.  This
satisfies 
\beq
N_{L_y,\lambda}=\sum_{d=0}^{L_y} n_P(L_y,d)
\label{npgen}
\eeq
where in the notation used here, $N_{L_y,\lambda}$ refers to the quantity
denoted $N_{P,L_y,\lambda}$ in \cite{cf}. 
We gave general formulas for $n_P(L_y,d)$ and $N_{L_y,\lambda}$; in
particular, here we shall need the following ones:
\beq
n_P(L_y,0)=M_{L_y-1}
\label{nplyd0}
\eeq
and
\beq
n_P(L_y,1)=M_{L_y}
\label{nply1m}
\eeq
where the Motzkin number $M_n$ is given by
\beq
M_n =  \sum_{j=0}^n (-1)^j C_{n+1-j} {n \choose j}
\label{motzkin}
\eeq
and
\beq
C_n=\frac{1}{n+1}{2n \choose n}
\label{catalan}
\eeq
is the Catalan number.  For the total number of terms, we obtained the result
\beq
N_{L_y,\lambda}=2(L_y-1)! \ \sum_{j=0}^{[\frac{L_y}{2}]} \frac{(L_y-j)}{
(j!)^2(L_y-2j)!} \ . 
\label{nptotform}
\eeq
Now our transformation formulas (\ref{cd0tran})-(\ref{cdoddtran}) imply that
the only cases where the coefficients in (\ref{pgsum}) 
for the cyclic and M\"obius strips can be
the same, up to sign, are for $c^{(0)}=1$ and $c^{(1)}$.  If these coefficients
are the same, then, by eq. (\ref{ccgminus}), $c_{cg,jm}=0$, while if they
are opposite in sign, then, by eq. (\ref{ccgplus}), $c_{cg,jp}=0$; hence, 
in
either case, there is only one term from the possible pair 
$\pm \lambda_{cg,sq,cyc,j}$ contributing to (\ref{pgsum}) for the cyclic
crossed-subgraph strip.  For all of the other coefficients $c_{sq,cyc,j}$ 
in (\ref{pgsum}) for the cyclic 
strips, our transformation formulas (\ref{cd0tran})-(\ref{cdoddtran}) imply
that $c_{sq,Mb,j} \ne \pm c_{sq,cyc,j}$, so that both $c_{cg,jp}$ and
$c_{cg,jm}$ are nonzero and both of the corresponding pair 
$\pm \lambda_{cg,sq,cyc,j}$ contribute to (\ref{pgsum}) for the cyclic
crossed-subgraph strip.  It follows that the total number of terms for the
cyclic crossed-subgraph strip of the square lattice is given by 
\beq
N_{cg,sq,L_y,\lambda} = 2N_{sq,cyc,L_y,\lambda}-n_P(L_y,0)-n_P(L_y,1) \ .
\label{ncgc0c1}
\eeq
Substituting our results from \cite{cf} for each of the quantities on the
right-hand side, we obtain, for the number of unpaired terms for the
crossed-subgraph strip, the relation 
\beq
N_{cg,sq,L_y,up,\lambda}=M_{L_y-1}+M_{L_y} \ , 
\label{ncgup}
\eeq
for the number of terms comprising members of opposite-sign pairs, the relation
\beqs
N_{cg,sq,L_y,opd,\lambda} & = & 2(N_{cg,sq,L_y,\lambda}-N_{sq,cyc,L_y,\lambda})
\cr\cr
& = & 2\biggl [ 2(L_y-1)! \ \sum_{j=0}^{[\frac{L_y}{2}]}
\frac{(L_y-j)}{(j!)^2(L_y-2j)!} - M_{L_y-1} - M_{L_y} \biggr ] 
\label{ncgopd}
\eeqs
and for the total number of (nonzero) terms, 
\beq
N_{cg,sq,L_y,\lambda} = 4(L_y-1)! \ \sum_{j=0}^{[\frac{L_y}{2}]} 
\frac{(L_y-j)}{(j!)^2(L_y-2j)!} - M_{L_y-1} - M_{L_y} \ . 
\label{ncgtot}
\eeq

\subsection{$L_y=2$ Cyclic Crossing-Subgraph Strip of the Square Lattice}

For example, for $L_y=2$, using the Motzkin numbers $M_1=1$, $M_2=2$, the
total number of terms is $N_{cg,sq,2,\lambda}=5$, with 
$N_{cg,sq,2,up,\lambda}=3$ and $N_{cg,sq,2,opd,\lambda}=2$.
We have explicitly calculated the chromatic polynomial for this case using 
iterated deletion-contraction and coloring matrix methods and find (with the
shorthand $cgs2$ for $cg(sq,2 \times L_x,cyc)$) 
\beq
P(cg(sq,2 \times L_x,cyc)) = \sum_{j=1}^5
c_{cgs2,j}(\lambda_{cgs2,j})^{L_x}
\label{pcgsq2}
\eeq
where
\beq
\lambda_{cgs2,j}=\pm 1 \ , j=1,2
\label{lamcgsq21}
\eeq
\beq
\lambda_{cgs2,3}=3-q
\label{lamcgsq23}
\eeq
\beq
\lambda_{cgs2,4}=q-1
\label{lamcgsq24}
\eeq
\beq
\lambda_{cgs2,5}=q^2-3q+3
\label{lamcgsq25}
\eeq
with coefficients
\beq
c_{cgs2,1}=\frac{1}{2}(c^{(2)}-c^{(0)}) =\frac{1}{2}q(q-3)
\label{ccgsq21}
\eeq
\beq
c_{cgs2,2}=\frac{1}{2}(c^{(2)}+c^{(0)}) =\frac{1}{2}(q-1)(q-2)
\label{ccgsq22}
\eeq
\beq
c_{cgs2,j}=c^{(1)}=q-1 \quad {\rm for} \ \ j=3,4
\label{ccgsq23}
\eeq
\beq
c_{cgs2,5}=c^{(0)}=1 \ . 
\label{cgsq25}
\eeq
For even $L_x$, this chromatic polynomial for the width $L_y=2$ crossing 
graph strip of the square lattice reduces to the result for the regular cyclic
strip of the square lattice \cite{bds} with 
$N_{sq,2,cyc,\lambda}=4$, namely, with $L_x=m$, 
\beq
P(sq,2 \times L_x, cyc.) = (q^2-3q+1) + (q-1)\biggl [ (3-q)^m + 
(1-q)^m \biggr ] + (q^2-3q+3)^m
\label{psqcyc2}
\eeq
while for odd $L_x$, the chromatic polynomial (\ref{pcgsq2}) reduces to the
result for the $L_y=2$ M\"obius strip with $N_{sq,2,Mb,\lambda}=4$, namely 
\cite{bds} 
\beq
P(sq,2 \times L_x, Mb) = -1 + (q-1)\biggl [ (3-q)^m -
(1-q)^m \biggr ] + (q^2-3q+3)^m \ . 
\label{psqmb2}
\eeq

\subsection{$L_y=3$ Cyclic Crossing-Subgraph Strip of the Square Lattice}

The chromatic polynomial for the $L_y=3$ cyclic 
crossing-subgraph strip of the square lattice (labelled $cgs3$)  
can be calculated directly or from the known results for the chromatic
polynomials of the $L_y=3$ cyclic \cite{wcy} and M\"obius \cite{pm} strips of
the square lattice.  It is worthwhile to display the results here because of
the unified understanding
that they give concerning the structures of the chromatic
polynomials for the cyclic and M\"obius strips.  From our general formulas
above we have $N_{cg,sq,3,\lambda}=14$ with $N_{cg,sq,3,up,\lambda}=6$ and 
$N_{cg,sq,3,opd,\lambda}=8$.  For the respective even and odd values of $L_x$
where the chromatic polynomial reduces to that for the $L_y=3$ cyclic and 
M\"obius strips of the square lattice, we have $N_{sq,3,cyc,\lambda}=
N_{sq,3,Mb,\lambda}=14-(1/2)*8=10$, in agreement with the previous calculations
in \cite{wcy,pm}.  We find 
\beq
P(cg(sq,3 \times L_x,cyc)) = \sum_{j=1}^{14} c_{cgs3,j}
(\lambda_{cgs3,j})^{L_x}
\label{pcgsq3}
\eeq
where
\beq
\lambda_{cgs3,j}=\pm 1 \ , j=1,2
\label{lamcgsq312}
\eeq
\beq
\lambda_{cgs3,j}=\pm (q-1) \ , j=3,4
\label{lamcgsq334}
\eeq
\beq
\lambda_{cgs3,j}=\pm (q-2) \ , j=5,6
\label{lamcgsq356}
\eeq
\beq
\lambda_{cgs3,j}=\pm (q-4) \ , j=7,8
\label{lamcgsq378}
\eeq
\beq
\lambda_{cgs3,9}=(q-2)^2
\label{lamcgsq39}
\eeq
\beq
\lambda_{cgs3,10}= \lambda_{sq3,6}
\label{lamcgsq310}
\eeq
\beq
\lambda_{cgs3,11}= \lambda_{sq3,7}
\label{lamcgsq311}
\eeq
where $\lambda_{sq3,j}$ for $j=6,7$ were given in eq. (3.10) of \cite{wcy}, 
and 
\beq
\lambda_{cgs3,j}=\lambda_{sq3,j-4} \ , 12 \le j \le 14
\label{lamcgsq31214}
\eeq
where $\lambda_{sq3,j}$ for $j=8,9,10$ were 
defined by eq. (3.11) of \cite{wcy}. 

The corresponding coefficients are 
\beq
c_{cgs3,1}=\frac{1}{2}(c^{(3)}-c^{(2)}) = \frac{1}{2}(q-2)(q^2-4q+1) 
\label{ccgsq31}
\eeq
\beq
c_{cgs3,2}=\frac{1}{2}(c^{(3)}+c^{(2)}) = \frac{1}{2}q(q-1)(q-3)
\label{ccgsq32}
\eeq
\beq
c_{cgs3,j}=\frac{1}{2}(c^{(2)}-c^{(0)})=\frac{1}{2}q(q-3) \quad {\rm for} \ \ 
j=3,6,7
\label{ccgsq33}
\eeq
\beq
c_{cgs3,j}=\frac{1}{2}(c^{(2)}+c^{(0)})=\frac{1}{2}(q-1)(q-2)
\quad {\rm for} \ \ j=4,5,8
\label{ccgsq34}
\eeq
\beq
c_{cgs3,j}=c^{(1)}=q-1 \quad {\rm for} \ \ j=9,12,13,14
\label{ccgsq39}
\eeq
\beq
c_{cgs3,j}=1 \ , j=10,11 \ . 
\label{ccgsq31011}
\eeq

\subsection{$L_y=4$ Cyclic Crossing-Subgraph Strip of the Square Lattice}

 From our calculations of the chromatic polynomials for the width $L_y=4$
cyclic and M\"obius strips of the square lattice \cite{s4}, we have obtained
the chromatic polynomial for the $L_y=4$ cyclic crossing-subgraph strip
(labelled $cgs4$).  In accord with our general formulas, we get 
$N_{cg,sq,4,\lambda}=39$ with $N_{cg,sq,4,up,\lambda}=13$ and 
$N_{cg,sq,4,opd,\lambda}=26$. 
For respective even and odd $L_x$, the reduction to
the $L_y=4$ cyclic and M\"obius strips has $N_{sq,4,cyc,\lambda}=
N_{sq,4,Mb,\lambda}=39-(1/2)*26=26$.  We omit listing the terms and their
coefficients here since they are lengthy and our previous examples are
sufficient to illustrate our general formulas.

In passing, we note that the cyclic crossing-subgraph strip of the triangular
lattice does not yield either the cyclic or M\"obius strip of this lattice for
even or odd $L_x$.  This provides a further understanding of our earlier
findings that the coefficients for the $L_y=2$ \cite{wcy} and $L_y=3$ \cite{t}
M\"obius strips of the triangular lattice are not polynomials in $q$.

\subsection{$L_y=3$ Toroidal Crossing-Subgraph Strip of the Square Lattice} 

For the toroidal crossing-subgraph strip of the square lattice with width 
$L_y=3$ (labelled $cgst3$) we find 
$N_{cgst3,\lambda}=12$ with $N_{cgst3,up,\lambda}=4$, 
$N_{cgst3,opd,\lambda}=2$, and $N_{cgst3,ops,\lambda}=6$ (see Table
\ref{proptable2}). Our result is 
\beq
P(cg(sq,3 \times L_x,PBC_y,PBC_x)) = \sum_{j=1}^{12} c_{cgst3,j}
(\lambda_{cgst3,j})^{L_x}
\label{pcgst3}
\eeq
with
\beq
\lambda_{cgst3,j}=\pm 1 \ , j=1,2
\label{lamcgst312}
\eeq
\beq
\lambda_{cgst3,3}=1-q
\label{lamcgsq33}
\eeq
\beq
\lambda_{cgst3,j} = \pm (q-2) \quad {\rm for} \ \ j=4,5
\label{lamcgsq345}
\eeq
\beq
\lambda_{cgst3,j} = \pm (q-4) \quad {\rm for} \ \ j=6,7
\label{lamcgsq367}
\eeq
\beq
\lambda_{cgst3,8}=q-5
\label{lamcgsq38}
\eeq
\beq
\lambda_{cgst3,j}=\pm (q-2)^2 \quad {\rm for} \ \ j=9,10
\label{lamcgsq3910}
\eeq
\beq
\lambda_{cgst3,11}=-(q^2-7q+13)
\label{lamcgsq3911}
\eeq
\beq
\lambda_{cgst3,12}=q^3-6q^2+14q-13
\label{lamcgsq3912}
\eeq
with corresponding coefficients
\beq
c_{cgst3,1}=\frac{1}{2}(q-2)(q^2-4q+1)
\label{ccgst31}
\eeq
\beq
c_{cgst3,2}=\frac{1}{2}q(q^2-6q+7)
\label{ccgst32}
\eeq
\beq
c_{cgst3,j}=\frac{1}{2}(q-1)(q-2) \quad {\rm for} \ \ j=3,6,7
\label{ccgst33}
\eeq
\beq
c_{cgst3,j}=\frac{1}{2}q(q-3) \quad {\rm for} \ \ j=4,5,8
\label{ccgst345}
\eeq
\beq
c_{cgst3,j}=q-1 \quad {\rm for} \ \ 9 \le j \le 11
\label{ccgst3910}
\eeq
\beq
c_{cgst3,12}=1 \ . 
\label{ccgst3912}
\eeq
Hence, in the even-$L_x$ case, the reduction to the
chromatic polynomial for the $L_y=3$ toroidal strip of the square lattice
(labelled $st3$) has, by eq. (\ref{nlamcgtorus}), $N_{st3,\lambda}=12-1-3=8$
terms, while in the odd-$L_x$ case, the reduction to the chromatic polynomial
for the $L_y=3$ Klein bottle strip of the square lattice (labelled $sk3$) has
$N_{sk3,\lambda}=12-1-6=5$.  These are in agreement with the results that were
obtained in \cite{tk}.  The resultant coefficients for the torus and Klein
bottle strips can be computed from the analogues of formulas 
(\ref{cgcoeffsum}) and
(\ref{cgcoeffdiff}) and agree with those given in \cite{tk}.

These correspondences shed further light on the coefficients that enter into
the chromatic polynomials for strip graphs with torus and Klein bottle boundary
conditions.  We recall that these are not of the simple form expressible in
terms of Chebyshev polynomials of the second kind that we found for strip
graphs with cyclic and M\"obius boundary conditions in \cite{cf}.  Among other
differences, there is not a unique coefficient with a given degree $d$ in $q$;
for example, for degree $d=2$, both $(q-1)(q-2)/2$ and $q(q-3)/2$ as
coefficients.  However, the reduction of the chromatic polynomials for the
toroidal crossing-subgraph strips to the respective chromatic polynomials for
the torus and Klein bottle strips yields relations linking the coefficients in
the latter to Chebyshev polynomials, such as (\ref{ccgsq21}) and
(\ref{ccgsq22}).  In the Appendix we list our results for the $L_y=4$
crossing-subgraph toroidal strip of the square lattice (labelled $cgst4$)

\begin{table}
\caption{\footnotesize{Numbers of nonzero terms $\lambda_{G_s,j}$, 
$N_{G_s,up,\lambda}$, $N_{G_s,opd,\lambda}$, and $N_{G_s,ops,\lambda}$
for crossing-subgraph strips of the square and triangular lattices.  
See text for notation.}}
\begin{center}
\begin{tabular}{|c|c|c|c|c|c|c|c|}
\hline\hline $G_s$ & $L_y$ & $BC_y$ & $BC_x$ & $N_{G_s,\lambda}$ & 
$N_{G_s,up,\lambda}$ & $N_{G_s,opd,\lambda}$ & $N_{G_s,ops,\lambda}$ \\ 
\hline\hline
cg,sq  & 3 & P & P   & 12 & 4  & 2  & 6   \\ \hline
cg,sq  & 4 & P & P   & 48 & 18 & 8  & 22  \\ \hline
cg,sq  & 2 & F & P   &  5 & 3  & 2  & 0   \\ \hline
cg,sq  & 3 & F & P   & 14 & 6  & 8  & 0   \\ \hline
cg,sq  & 4 & F & P   & 39 & 13 & 26 & 0   \\ \hline
cg,tri & 3 & P & P   & 12 & 4  & 2  & 6   \\ \hline
cg,tri & 4 & P & P   & 40 & 10 & 4  & 26  \\ \hline
\end{tabular}
\end{center}
\label{proptable2}
\end{table}

\section{Toroidal Crossing-Subgraph Strips of the Triangular Lattice} 

\begin{figure}[h]
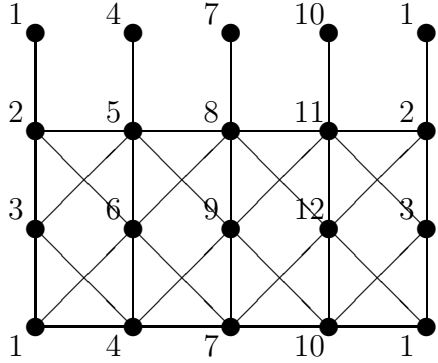

\caption{\footnotesize{Illustrative toroidal crossing-subgraph strip graph 
of the triangular lattice with $L_y=3$ and $L_x=4$. Vertices are indicated with
$\bullet$ (points where edges cross without a symbol $\bullet$ are not
vertices.)}}
\label{cgraphtri}
\end{figure}

We consider here a strip of the triangular lattice of fixed width $L_y$ and
arbitrarily great length $L_x$ constructed as follows. As before, label the
vertices of two successive transverse slices of the strip, starting at the top,
as $(1,2,..,L_y)$ and $(1^\prime,2^\prime,...,L_y^\prime)$.  With periodic
transverse boundary conditions, these transverse slices form circuit graphs,
$C_{L_y}$.  Connect these with edges linking vertices 1 to $L_y^\prime$, 2 to
$(L_y-1)^\prime$, and so forth.  This forms the toroidal crossing-subgraph
strip of the square lattice.  Next, add diagonal edges as illustrated for the
$L_y=3$ case in Fig. \ref{cgraphtri}.  Finally, impose periodic longitudinal
boundary conditions.  This yields the crossing-subgraph (cg) strip of the
triangular (tri) lattice with toroidal boundary conditions. We shall label this
strip as $cg(tri,L_y \times L_x, PBC_y,PBC_x)$. 

\vspace*{5mm}
\begin{center}
\begin{picture}(40,30)
\multiput(0,0)(10,0){5}{\circle*{2}}
\multiput(0,10)(10,0){5}{\circle*{2}}
\multiput(0,20)(10,0){5}{\circle*{2}}
\multiput(0,30)(10,0){5}{\circle*{2}}
\multiput(0,0)(10,0){5}{\line(0,1){30}}
\put(0,0){\line(1,0){40}}
\put(0,20){\line(1,0){40}}
\multiput(0,0)(10,0){4}{\line(1,1){10}}
\multiput(0,10)(10,0){4}{\line(1,-1){10}}
\multiput(0,10)(10,0){4}{\line(1,1){10}}
\multiput(0,20)(10,0){4}{\line(1,-1){10}}
\put(-2,-2){\makebox(0,0){1}}
\put(8,-2){\makebox(0,0){4}}
\put(18,-2){\makebox(0,0){7}}
\put(28,-2){\makebox(0,0){10}}
\put(38,-2){\makebox(0,0){1}}
\put(-2,12){\makebox(0,0){3}}
\put(8,12){\makebox(0,0){6}}
\put(18,12){\makebox(0,0){9}}
\put(28,12){\makebox(0,0){12}}
\put(38,12){\makebox(0,0){3}}
\put(-2,22){\makebox(0,0){2}}
\put(8,22){\makebox(0,0){5}}
\put(18,22){\makebox(0,0){8}}
\put(28,22){\makebox(0,0){11}}
\put(38,22){\makebox(0,0){2}}
\put(-2,32){\makebox(0,0){1}}
\put(8,32){\makebox(0,0){4}}
\put(18,32){\makebox(0,0){7}}
\put(28,32){\makebox(0,0){10}}
\put(38,32){\makebox(0,0){1}}
\end{picture}
\end{center}

\noindent 

We  first observe that
\beq
cg(tri,L_y \times L_x,PBC_y,PBC_x)= tri(L_y \times L_x, PBC_y,PBC_x) \quad 
{\rm if} \ \ L_x = 0 \ {\rm mod} \ 2L_y
\label{cgtritorusidentity}
\eeq
i.e., for $L_x=0$ mod $2L_y$, the toroidal crossing-subgraph strip of the
triangular lattice reduces to the toroidal strip of this lattice.  Associated
with this result, there are terms in the chromatic polynomial for the 
resultant toroidal strip of the triangular lattice corresponding to 
opposite-sign pairs of $\lambda_{cgtL_y,j}$'s in the chromatic polynomial 
for the toroidal crossing-subgraph strip that appear with phase factors of the
form $e^{\pm \pi i/L_y}$ and certain products thereof.
For $L_y=3$ these are $e^{\pm 2\pi i/3}$, while for $L_y=4$ they are 
$e^{\pm \pi \ell i/4}$, $\ell=1,2$.  It is thus necessary to distinguish 
between terms $\lambda_{cgtL_y,j}$ in the chromatic polynomial for the $L_y$ 
toroidal 
crossing-subgraph strip of the triangular lattice that correspond to terms in
the toroidal strip of the triangular lattice that do, or do not, have the form
of complex-conjugate pairs of terms $\lambda_{ttL_y,j}$ with complex
prefactors such as $e^{\pm 2\pi i/3}$. (where the subscript
$ttL_y$ refers to the toroidal strip of the triangular lattice with width
$L_y$).  We thus define $N_{cg,tri,L_y,torus,ops,r,\lambda}$ and 
$N_{cg,tri,L_y,torus,ops,i,\lambda}$ 
respectively as the number of $\lambda_{cgtL_y,j}$'s
that comprise opposite-sign pairs such that the members of each pair have the
same coefficient and correspond to a $\lambda_{ttL_y,j}$ that is real (is a
member of a pair with complex prefactor) for real $q$.  The resultant general
formula relating these numbers of terms is
\beq
N_{tri,L_y,torus,\lambda}=N_{cg,tri,L_y,torus,\lambda}
-\frac{1}{2}N_{cg,tri,L_y,torus,opd,\lambda}
-\frac{1}{2}N_{cg,tri,L_y,torus,ops,r,\lambda}
\label{nlamcgtritorus}
\eeq

If (i) $L_y$ is odd and $L_x$ is odd, or (ii) if $L_y$ is even and $L_x=1$ mod
4, then the toroidal crossing-subgraph strip of the triangular lattice reduces
to the strip of this lattice with Klein bottle boundary conditions
$(PBC_y,TPBC_x)$.  In this case, the reduction of the number of terms is
determined by eq. (\ref{nlamcgkb}) with the obvious replacement of square by
triangular lattice strip, i.e., 
\beq
N_{tri,L_y,Kb,\lambda}=N_{cg,tri,L_y,torus,\lambda}-
\frac{1}{2}N_{cg,tri,L_y,torus,opd,\lambda}-
N_{cg,tri,L_y,torus,ops,\lambda} \ . 
\label{nlamcgtrikb}
\eeq
Thus, 
\beq
N_{tri,L_y,torus,\lambda}-N_{tri,L_y,Kb,\lambda}=
N_{cg,tri,L_y,torus,ops,\lambda} 
-\frac{1}{2}N_{cg,tri,L_y,torus,ops,r,\lambda} \ .
\label{nlamcgtoruskbtri}
\eeq

In the limit $L_x \to \infty$, the width $L_y$ toroidal crossing-subgraph strip
of the triangular lattice yields a $W$ function via (\ref{w}) and hence a
singular locus ${\cal B}$.  As before, we can take this limit using values of
$L_x$ such that the crossing-subgraph strip reduces to the strip of the
triangular lattice with either torus or Klein bottle boundary conditions.
Given that the original limit exists, all three of these limits must be the
same.  This proves

\vspace{4mm}

\begin{flushleft} 

Theorem 2: \ The $W$ function and singular locus ${\cal B}$ are the same for
the $L_x \to \infty$ limit of strip of the triangular lattice with width $L_y$
and length $L_x$ whether one imposes $(PBC_y,PBC_x)$ or $(PBC_y,TPBC_x)$, i.e.
torus or Klein bottle boundary conditions.

\vspace{4mm}

Hence also

\vspace{4mm}

Corollary 2: \ Consider an $L_y \times L_x$ strip of the triangular lattice
with torus or Klein bottle boundary conditions, and denote the set of nonzero
eigenvalues that contribute to (\ref{pgsum}) for these two respective strips as
$\lambda_{ttL_y,j}$, $j=1,..,N_{ttL_y,\lambda}$ and $\lambda_{tkL_y,j}$,
$j=1,..,N_{tkL_y,\lambda}$. Focus on the set of eigenvalues $\lambda_{ttL_y,j}$
that do not occur among the set $\lambda_{tkL_y,j}$ (the number of these is
given by (\ref{nlamcgtoruskbtri})); none of these can be dominant eigenvalues.

\end{flushleft}
\vspace{4mm}

For the $L_y=3$ crossing-subgraph toroidal strip of the triangular lattice
(labelled $cgt3$) we find that $N_{cgt3,\lambda}=12$ and calculate
\beq
P(cg(tri,3 \times L_x,torus)) = \sum_{j=1}^{12} c_{cgt3,j}
(\lambda_{cgt3,j})^{L_x}
\label{pcgtri3}
\eeq
where
\beq
\lambda_{cgt3,j}=\pm 1 \quad {\rm for} \ \ j=1,2
\label{lamcgt312}
\eeq
\beq
\lambda_{cgt3,j}=\pm 2 \quad {\rm for} \ \ j=3,4
\label{lamcgt334}
\eeq
\beq
\lambda_{cgt3,j}=\pm (2q-7) \quad {\rm for} \ \ j=5,6
\label{lamcgt356}
\eeq
\beq
\lambda_{cgt3,7}=2-q
\label{lamcgt37}
\eeq
\beq
\lambda_{cgt3,8}=3q-14
\label{lamcgt38}
\eeq
\beq
\lambda_{cgt3,9}=-2(q-4)^2
\label{lamcgt39}
\eeq
\beq
\lambda_{cgt3,j}=\pm (q^2-5q+7) \quad {\rm for} \ \ j=10,11
\label{lamcgt31011}
\eeq
\beq
\lambda_{cgt3,12}=q^3-9q^2+29q-32 \ .
\label{lamcgt3_12}
\eeq
The corresponding coefficients are
\beq
c_{cgt3,j}=\frac{1}{2}q(q-1)(2q-7) \quad {\rm for} \ \ j=1,2
\label{ccgt3_12}
\eeq
\beq
c_{cgt3,3}=\frac{1}{6}(q-1)(q-2)(q-3)
\label{ccgt33}
\eeq
\beq
c_{cgt3,4}=\frac{1}{6}q(q-1)(q-5)
\label{ccgt34}
\eeq
\beq
c_{cgt3,j}=\frac{1}{2}(q-1)(q-2) \quad {\rm for} \ \ 5 \le j \le 7
\label{ccgt357}
\eeq
\beq
c_{cgt3,8}=\frac{1}{2}q(q-3)
\label{ccgt358}
\eeq
\beq
c_{cgt3,j}=q-1 \quad {\rm for} \ \ 9 \le j \le 11
\label{ccgt911}
\eeq
\beq
c_{cgt3,12}=1 \ . 
\label{ccgt312}
\eeq 
Thus, $N_{cgt3,up,\lambda}=4$, $N_{cgt3,opd,\lambda}=2$, and
$N_{cgt3,ops,\lambda}=6$.  Hence, for $L_x=0$ mod 6, where the chromatic
polynomial reduces to that for the $L_y=3$ toroidal strip of the triangular
lattice, the number of $\lambda_j$'s is reduced, according to the general
formula (\ref{nlamcgtritorus}), to $N_{tt3,\lambda}=12-1=11$. 
These numbers agree with our previous result \cite{t}, 
\beq
P(tri,3 \times L_x,torus) = \sum_{j=1}^{11} c_{tt3,j}
(\lambda_{tt3,j})^{L_x}
\label{ptritor3}
\eeq
where
\beq
\lambda_{tt3,1}=-2 
\label{lamtt1}
\eeq
\beq
\lambda_{tt3,2}=q-2
\label{lamtt2}
\eeq
\beq
\lambda_{tt3,3}=3q-14
\label{lamtt3}
\eeq
\beq
\lambda_{tt3,4}=-2(q-4)^2
\label{lamtt4}
\eeq
\beq
\lambda_{tt3,5}=q^3-9q^2+29q-32
\label{lamtt5}
\eeq
\beq
\lambda_{tt3,j}=e^{\pm 2\pi i/3} \quad {\rm for} \ \ j=6,7
\label{lamtt67}
\eeq
\beq
\lambda_{tt3,j}=(q^2-5q+7)e^{\pm 2\pi i/3}  \quad {\rm for} \ \ j=8,9
\label{lamtt89}
\eeq
\beq
\lambda_{tt3,j}=-(2q-7)e^{\pm 2\pi i/3}  \quad {\rm for} \ \ j=10,11
\label{lamtt1011}
\eeq
with coefficients 
\beq
c_{tt3,1} = \frac{1}{3}(q-1)(q^2-5q+3)
\label{ctt31}
\eeq
\beq
c_{tt3,j}=\frac{1}{2}(q-1)(q-2) \quad {\rm for} \ \ j=2,10,11
\label{ctt32}
\eeq
\beq
c_{tt3,3}=\frac{1}{2}q(q-3)
\label{ctt33}
\eeq
\beq
c_{tt3,j}=q-1 \quad {\rm for} \ \ j=4,8,9
\label{ctt34}
\eeq
\beq
c_{tt3,5}=1
\label{ctt35}
\eeq
\beq
c_{tt3,6} = c_{tt3,7} = \frac{1}{6}q(q-1)(2q-7) \ . 
\label{ctt367}
\eeq

Similarly, for odd $L_x$
where the $L_y=3$ crossing-subgraph strip reduces to the $L_y=3$ Klein bottle
strip of the triangular lattice, the number of nonzero terms is reduced,
according to the general formula (\ref{nlamcgtrikb}), to $N_{tk3,\lambda}=
12-1-6=5$, in agreement with our previous calculation
\cite{t}
\beq
P(tri,3 \times L_x,PBC_y,TPBC_x) = \sum_{j=1}^5 c_{tk3,j}
(\lambda_{tk3,j})^{L_x}
\label{ptrik3}
\eeq
where
\beq
\lambda_{tk3,1}=\lambda_{tt3,1}=-2
\label{lamtk31}
\eeq
\beq
\lambda_{tk3,2}=\lambda_{tt3,2}=q-2
\label{lamtk32}
\eeq
\beq
\lambda_{tk3,3}=\lambda_{tt3,3}=3q-14
\label{lamtk33}
\eeq
\beq
\lambda_{tk3,4}=\lambda_{tt3,4}=-2(q-4)^2
\label{lamtk34}
\eeq
\beq
\lambda_{tk3,5}=\lambda_{tt3,5}=q^3-9q^2+29q-32
\label{lamtk35}
\eeq
with coefficients
\beq
c_{tk3,1}=-(q-1)
\label{ctk31}
\eeq
\beq
c_{tk3,2}=-\frac{1}{2}(q-1)(q-2)
\label{ctk32}
\eeq
\beq
c_{tk3,3}=\frac{1}{2}q(q-3)
\label{ctk33}
\eeq
\beq
c_{tk3,4}=q-1
\label{ctk34}
\eeq
\beq
c_{tk3,5}=1 \ . 
\label{ctk35}
\eeq 
In the Appendix we give our results for the $L_y=4$ crossing-subgraph toroidal
strip of the triangular lattice.

\section{Concluding Discussion}

We comment further here on some features of our results. 

\begin{enumerate}

\item 

Our exact calculations of the singular loci ${\cal B}$ for $L_y=4$ strips of
the square and triangular lattice with toroidal or Klein bottle conditions
exhibit the following features, as did the earlier calculations for the $L_y=3$
strips of these lattices in \cite{tk,t}: ${\cal B}$ (i) passes through $q=0$,
(ii) passes through $q=2$, (iii) passes through a maximal real point, thereby
defining a $q_c$, and (iv) encloses one or more regions including the interval
$0 < q < q_c$ \cite{w}.  As noted above, we also found that these four features
hold for the ($L_x \to \infty$ limit of) strips with cyclic and M\"obius
boundary conditions, which leads to the inference that the key condition is the
existence of periodic (or reversed-orientation periodic) longitudinal boundary
conditions. 

\item 

Previous exact calculations of ${\cal B}$ for cyclic and M\"obius strips of the
square and triangular lattice of various widths \cite{w,wcy,pm,t,s4} are
consistent with the inference that as $L_y$ increases, the outer envelope of
${\cal B}$ moves outward, i.e. if $L_y > L_y^\prime$, then the outer envelope
of ${\cal B}$ for $L_y$ encloses that for $L_y^\prime$ \cite{bcc}.  In
particular, $q_c$ is a nondecreasing function of $L_y$.  However, we have also
shown that neither of these properties holds for strips with $(PBC_y,FBC_x)$,
i.e., cylindrical, boundary conditions \cite{strip2,t,s4}.  Our present results
show that for strips of the square and triangular lattices with toroidal or
Klein bottle boundary conditions, $(PBC_y,(T)PBC_x)$, the outer envelope of
${\cal B}$ does not, in general, move monotonically outward as one increases
the width.  This is illustrated in Figs. \ref{sqpxpy34} and \ref{tpxpy34},
which show the boundaries ${\cal B}$ for the $L_y=3$ and $L_y=4$ strips of,
respectively, the square and triangular lattices with torus or Klein bottle
boundary conditions.  See also Table \ref{proptable}.  This monotonic
(nonmonotonic) behavior of the outer envelope is reminiscent of the monotonic
(nonmonotonic) behavior of the $W$ function for free (periodic) transverse
boundary conditions discussed in \cite{w2d} (see also \cite{t,s4}).

\begin{figure}[hbtp]
\centering
\leavevmode
\epsfxsize=3.0in
\begin{center}
\leavevmode
\epsffile{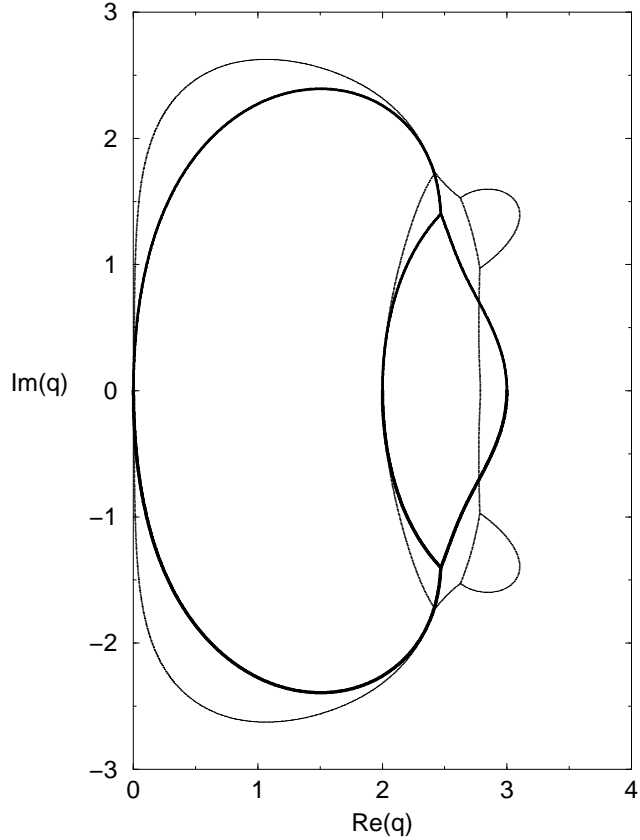}
\end{center}
\caption{\footnotesize{Comparison of the singular loci ${\cal B}$ for the $L_x
\to \infty$ limit of the strips of the square lattice with $L_y=3$ (darker
curve) and $L_y=4$ (lighter curve) with toroidal boundary conditions (or
equivalently, Klein bottle boundary conditions, which yield the same ${\cal B}$
for a given $L_y$).}}
\label{sqpxpy34}
\end{figure}

\begin{figure}[hbtp]
\centering
\leavevmode
\epsfxsize=3.0in
\begin{center}
\leavevmode
\epsffile{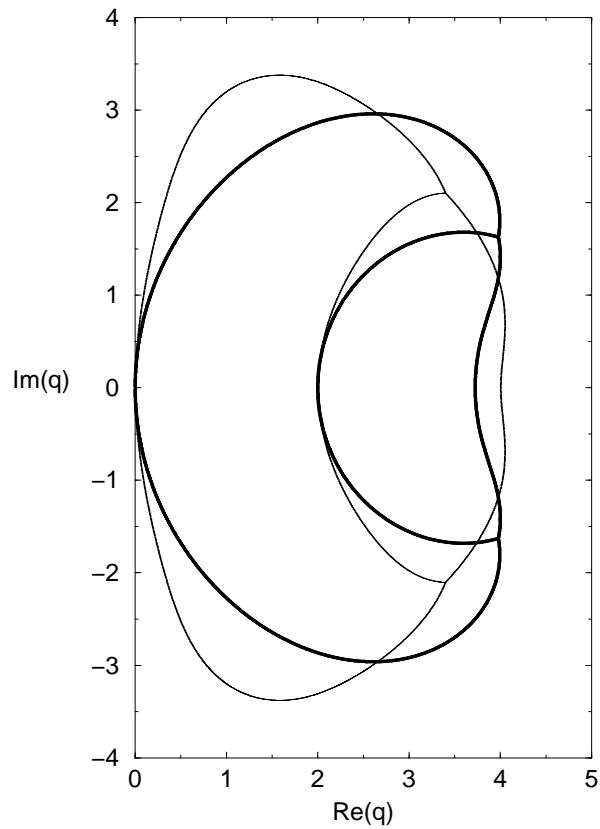}
\end{center}
\caption{\footnotesize{Comparison of the singular loci ${\cal B}$ for the $L_x
\to \infty$ limit of the strips of the triangular lattice with $L_y=3$ (darker
curve) and $L_y=4$ (lighter curve) with toroidal boundary conditions (or
equivalently, Klein bottle boundary conditions, which yield the same ${\cal B}$
for a given $L_y$).}}
\label{tpxpy34}
\end{figure}

\item 

As a special aspect of this outer envelope, $q_c$ decreases from 3 to
approximately 2.78 for the (infinite-length limit of the) square-lattice strip
with toroidal or Klein bottle boundary conditions when one increase the width
from $L_y=3$ to $L_y=4$.  In contrast, $q_c$ increases from about 3.72 to 4 for
the (infinite-length limit of the) triangular-lattice strip with toroidal or
Klein bottle boundary conditions when one increases $L_y$ from 3 to 4.  Related
to this, the calculation of $P$ and $W$ and ${\cal B}$ for the $L_y=3$ strip of
the square lattice with toroidal or Klein bottle boundary conditions in
\cite{tk} showed that $q_c=3$ for (the $L_x \to \infty$ limit of) that strip,
and hence showed that $q_c$ for a finite-width, infinite-length strip of a
given lattice can be the same as for the limit of infinite width, i.e. the full
2D infinite lattice, since $q_c=3$ for the square lattice \cite{lieb} (for
general upper bounds, see \cite{ssbounds}).  So far, this was an isolated
example.  Our calculation in the present paper provides a second example of
this phenomenon: $q_c$ for the infinite-length limit of the $L_y=4$ strip of
the triangular lattice with toroidal or Klein bottle boundary conditions has
the value $q_c=4$, which is equal to the value \cite{baxter} for the full 2D
triangular lattice, i.e. the $L_y \to \infty$ limit of the strip.
Parenthetically, we note that rigorous bounds on $q_c$ have been given in
\cite{ssbounds}.

\item 

In all of the cases of strips of the square and triangular lattice with
periodic or reversed-orientation periodic longitudinal boundary conditions for
which we have performed exact calculations of the chromatic polynomials and
have determined the respective singular loci ${\cal B}$, we have found the
following results for the coefficients corresponding to the dominant terms in
various regions: (i) in region $R_1$, this coefficient has been proved to be
unity \cite{bcc}; (ii) in the region containing the interval $0 < q < 2$, the
coefficient is $c^{(1)}=q-1$ for the strips of the square and triangular
lattice with cyclic and torus b.c., the square strips with M\"obius and Klein
bottle b.c. and the triangular strips with Klein bottle b.c. (the coefficients
are not, in general, polynomials for M\"obius strips of the triangular
lattice); and (iii) for the observed complex-conjugate pairs of regions, the
coefficients are also $q-1$ for cyclic and torus b.c. and $\pm (q-1)$ for
M\"obius ($sq$ case) and Klein bottle b.c. \cite{w,wcyl,wcy,pm,t,s4}.  A fourth
finding is that (iv) for the torus/Klein bottle boundary conditions, in the
cases that we have studied, we have found that the coefficient corresponding to
the dominant $\lambda_{G,j}$ in the region containing the interval $2 < q <
q_c$, for each respective $q_c$, is $q(q-3)/2$.

\end{enumerate}

Acknowledgment: The research of R. S. was supported in part by the NSF
grant PHY-9722101 and at Brookhaven by the DOE contract
DE-AC02-98CH10886.\footnote{\footnotesize{Accordingly, the U.S. government
retains a non-exclusive royalty-free license to publish or reproduce the
published form of this contribution or to allow others to do so for
U.S. government purposes.}}

\section{Appendix: Further Results on Crossing-Subgraph Strips}

\subsection{$L_y=4$ Toroidal Crossing-Subgraph Strip of the Square Lattice}

For the $L_y=4$ crossing-subgraph 
toroidal strip of the square lattice (labelled
$cgst4$) we find that $N_{cgst4,\lambda}=48$ and calculate 
\beq
P(cg(sq,4 \times L_x,torus)) = \sum_{j=1}^{48} c_{cgst4,j}
(\lambda_{cgst4,j})^{L_x}
\label{pcgsq4}
\eeq
where
\beq
\lambda_{cgst4,j}=\pm 1 \quad {\rm for} \ \ j=1,2
\label{lamcgs412}
\eeq
\beq
\lambda_{cgst4,j}=\pm (q-1) \quad {\rm for} \ \ j=3,4
\label{lamcgs434}
\eeq
\beq
\lambda_{cgst4,j}=\pm (q-2) \quad {\rm for} \ \ j=5,6
\label{lamcgs456}
\eeq
\beq
\lambda_{cgst4,j}=\pm (q-3) \quad {\rm for} \ \ j=7,8
\label{lamcgs478}
\eeq
\beq
\lambda_{cgst4,j}=\pm (q-4) \quad {\rm for} \ \ j=9,10
\label{lamcgs4910}
\eeq
\beq
\lambda_{cgst4,j}=\pm (q-5) \quad {\rm for} \ \ j=11,12
\label{lamcgs4910a}
\eeq
\beq
\lambda_{cgst4,13}=-(q^2-5q+5)
\label{lamcgs413}
\eeq
\beq
\lambda_{cgst4,14}=q^2-5q+7
\label{lamcgs414}
\eeq
\beq
\lambda_{cgst4,15}=-(q-1)(q-3)
\label{lamcgs415}
\eeq
\beq
\lambda_{cgst4,j}=\lambda_{st4,j-6} \quad {\rm for} \ \ j=16,17
\label{lamcgs41617}
\eeq
(where $\lambda_{st4,j}$ for $j=10,11$ were given in the text in 
eq. (\ref{lamtorsq1011})), 
\beq
\lambda_{cgst4,j}=\pm \lambda_{st4,12} \quad {\rm for} \ \ j=18,19
\label{lamcgs41819}
\eeq
\beq
\lambda_{cgst4,j}=\pm \lambda_{st4,13} \quad {\rm for} \ \ j=20,21
\label{lamcgs42021}
\eeq
\beq
\lambda_{cgst4,j}=- \lambda_{st4,j-8} \quad {\rm for} \ \ j=22,23
\label{lamcgs42223}
\eeq
\beq
\lambda_{cgst4,j}=\pm \lambda_{st4,16} \quad {\rm for} \ \ j=24,25
\label{lamcgs42425}
\eeq
\beq
\lambda_{cgst4,j}=\pm \lambda_{st4,17} \quad {\rm for} \ \ j=26,27
\label{lamcgs42627}
\eeq
\beq
\lambda_{cgst4,j}=-\lambda_{st4,j-10} \quad {\rm for} \ \ j=28,29
\label{lamcgs42829}
\eeq
\beq
\lambda_{cgst4,j}=\pm \lambda_{st4,20} \quad {\rm for} \ \ j=30,31
\label{lamcgs43031}
\eeq
\beq
\lambda_{cgst4,j}=\pm \lambda_{st4,21} \quad {\rm for} \ \ j=32,33
\label{lamcgs43233}
\eeq
\beq
\lambda_{cgst4,j}=-\lambda_{st4,j-12} \quad {\rm for} \ \ 34 \le j \le 36
\label{lamcgs43436}
\eeq
\beq
\lambda_{cgst4,j}=\lambda_{st4,j-12} \quad {\rm for} \ \ 37 \le j \le 42
\label{lamcgs43742}
\eeq
\beq
\lambda_{cgst4,j}=\pm \lambda_{st4,31} \quad {\rm for} \ \ j=43,44
\label{lamcgs44344}
\eeq
\beq
\lambda_{cgst4,j}=\pm \lambda_{st4,32} \quad {\rm for} \ \ j=45,46
\label{lamcgs4546}
\eeq
\beq
\lambda_{cgst4,j}=\pm \lambda_{st4,33} \quad {\rm for} \ \ j=47,48 \ . 
\label{lamcgs4748}
\eeq

The corresponding coefficients are
\beq
c_{cgst4,1}=\frac{1}{2}(q-1)(q^3-7q^2+13q-2)
\label{ccgs41}
\eeq
\beq
c_{cgst4,2}=\frac{1}{2}q(q-3)(q^2-5q+5)
\label{ccgs42}
\eeq
\beq
c_{cgst4,j}=\frac{1}{6}q(q-1)(q-5) \quad {\rm for} \ \ j=3,12
\label{ccgs43}
\eeq
\beq
c_{cgst4,4}=\frac{1}{6}(q-1)(q-2)(q-3)
\label{ccgs44}
\eeq
\beq
c_{cgst4,j}=\frac{1}{3}q(q-2)(q-4) \quad {\rm for} \ \ j=5,9,10 \ \ {\rm and} \
\ 30 \le j \le 33
\label{ccgs45}
\eeq
\beq
c_{cgst4,6}=\frac{1}{6}(q-2)(q-3)(2q+1)
\label{ccgs46}
\eeq
\beq
c_{cgst4,j}=\frac{1}{3}(q-1)(q^2-5q+3) \quad {\rm for} \ \ j=7,8
\label{ccgs478}
\eeq
\beq
c_{cgst4,11}=\frac{1}{6}(q-1)(q-2)(q-3)
\label{ccgs411}
\eeq
\beq
c_{cgst4,j}=\frac{1}{2}(q-1)(q-2) \quad {\rm for} \ \ j=13,28,29 \ \ 
{\rm and} \ \ 43 \le j \le 48
\label{ccgs413}
\eeq
\beq
c_{cgst4,j}=\frac{1}{2}q(q-3) \quad {\rm for} \ \ j=14 \ \ {\rm and} \ \ 
22 \le j \le 27, \ \ 40 \le j 42
\label{ccgs414}
\eeq
\beq
c_{cgst4,j}=1 \quad {\rm for} \ \ j=15,16,17
\label{ccgs41517}
\eeq
\beq
c_{cgst4,j}=q-1 \quad {\rm for} \ \ 18 \le j \le 21 \ \ {\rm and} \ \ 
34 \le j \le 39 \ . 
\label{ccgs41821}
\eeq
Thus, $N_{cgst4,opd,\lambda}=8$ and $N_{cgst4,ops,\lambda}=22$ and hence for
the even and odd $L_x$ values where the chromatic polynomial for this
crossing-subgraph strip reduces to the respective chromatic polynomial for the
$L_y=4$ strip with torus and Klein bottle boundary conditions, we have, from
eqs. (\ref{nlamcgtorus}) and (\ref{nlamcgkb}),
$N_{st4,\lambda}=48-(1/2)*(8+22)=33$ and $N_{sk4,\lambda}=48-(1/2)*8-22=22$, in
agreement with our calculations in the text.

\subsection{$L_y=4$ Toroidal Crossing-Subgraph Strip of the Triangular
Lattice}

For the $L_y=4$ crossing-subgraph toroidal strip of the triangular lattice 
(labelled $cgt4$) 
we find that there are $N_{cgt4,\lambda}=40$ different nonzero 
$\lambda_{cgt4,j}$ terms that enter into the chromatic polynomial and that 
the coloring matrix also has a zero eigenvalue, so that the total number of
eigenvalues of the coloring matrix for this strip is $N_{cgt4,\lambda,tot}=41$.
We calculate 
\beq
P(cg(tri,4 \times L_x,torus)) = \sum_{j=1}^{40} c_{cgt4,j}
(\lambda_{cgt4,j})^{L_x}
\label{pcgtri4}
\eeq
where 
\beq
\lambda_{cgt4,j}=\pm 2 \quad {\rm for} \ \ j=1,2
\label{lamcgt412}
\eeq
\beq
\lambda_{cgt4,j}=\pm \sqrt{2} \quad {\rm for} \ \ j=3,4
\label{lamcgt434}
\eeq
\beq
\lambda_{cgt4,5}=2(3-q)
\label{lamcgt45}
\eeq
\beq
\lambda_{cgt4,j}=\pm (3-q) \quad {\rm for} \ \ j=6,7
\label{lamcgt467}
\eeq
\beq
\lambda_{cgt4,j}=\pm 2(2q-9)  \quad {\rm for} \ \ j=8,9
\label{lamcgt489}
\eeq
\beq
\lambda_{cgt4,10}=-2(q-3)^2
\label{lamcgt410}
\eeq
\beq
\lambda_{cgt4,j}=\lambda_{tt4,j-3} \quad {\rm for} \ \ j=11,12
\label{lamcgt41112}
\eeq
where $\lambda_{tt4,j}$ for $j=8,9$ were given above in
eqs. (\ref{lamtritor89}),
\beq
\lambda_{cgt4,j}=\pm \sqrt{3}(q-3) \quad {\rm for} \ \ j=13,14
\label{lamcgt41314}
\eeq
\beq
\lambda_{cgt4,j}=\pm (q-2)\sqrt{2(q-3)(q-4)} \quad {\rm for} \ \ j=15,16
\label{lamcgt41516}
\eeq
\beq
\lambda_{cgt4,j}=\lambda_{tt4,j-3} \quad {\rm for} \ \ 17 \le j \le 22 \ .
\label{lamcgt41722}
\eeq
The twelve terms $\lambda_{cgt4,j}$ for $23 \le j \le 34$ are related to
the
$\lambda_{tt4,j}$'s that are the roots of the quartic equations
(\ref{eqquartic1tritor})-(\ref{eqquartic3tritor}) as follows, where the 
$a_\ell$ and $b_\ell$ were
defined in eqs. (\ref{lamtritor2021})-(\ref{lamtritor3031}):
\beq
\lambda_{cgt4,j}=\pm \sqrt{2} a_1 \quad {\rm for} \ \ j=23,24
\label{lamcgt42324}
\eeq
\beq
\lambda_{cgt4,j}=\pm \sqrt{2} b_1 \quad {\rm for} \ \ j=25,26
\label{lamcgt42526}
\eeq
\beq
\lambda_{cgt4,j}=\pm \sqrt{2} a_2 \quad {\rm for} \ \ j=27,28
\label{lamcgt42728}
\eeq
\beq
\lambda_{cgt4,j}=\pm \sqrt{2} b_2 \quad {\rm for} \ \ j=29,30
\label{lamcgt42930}
\eeq
\beq
\lambda_{cgt4,j}=\pm \sqrt{2} a_3 \quad {\rm for} \ \ j=31,32
\label{lamcgt43132}
\eeq
\beq
\lambda_{cgt4,j}=\pm \sqrt{2} b_3 \quad {\rm for} \ \ j=33,34 \ . 
\label{lamcgt43334}
\eeq
The six terms $\lambda_{cgt4,j}$ for $35 \le j \le 40$ are related to the
$\lambda_{tt4,j}$'s that are the roots of the sixth-degree equation
(\ref{eqsixtritor}) as follows, where the $c_\ell$ were
defined in eqs. (\ref{lamtritor3233})-(\ref{lamtritor3637}):
\beq
\lambda_{cgt4,j}=\pm \sqrt{2} c_1 \quad {\rm for} \ \ j=35,36
\label{lamcgt43536}
\eeq
\beq
\lambda_{cgt4,j}=\pm \sqrt{2} c_2 \quad {\rm for} \ \ j=37,38
\label{lamcgt43738}
\eeq
\beq
\lambda_{cgt4,j}=\pm \sqrt{2} c_3 \quad {\rm for} \ \ j=39,40 \ .
\label{lamcgt43940}
\eeq

The corresponding coefficients are
\beq
c_{cgt4,1}=\frac{1}{8}q(q-1)(q-2)(q-3)
\label{ccgt4_1}
\eeq
\beq
c_{cgt4,2}=\frac{1}{8}q(q-2)(q-3)(q-5)
\label{ccgt4_2}
\eeq
\beq
c_{cgt4,j}=\frac{1}{12}(q-1)(q-2)(3q^2-11q-6) \quad {\rm for} \ \ j=3,4
\label{ccgt4_45}
\eeq
\beq
c_{cgt4,j}=\frac{1}{2}(q-1)(q-2) \quad {\rm for} \ \ j=5,10 \ \ {\rm and} \ \ 
35 \le j \le 40
\label{ccgt4_5}
\eeq
\beq
c_{cgt4,j}=\frac{1}{3}q(q-2)(q-4) \quad {\rm for} \ \ j=6,7,13,14 \ \ {\rm and}
\ \ 31 \le j \le 34
\label{ccgt4_67}
\eeq
\beq
c_{cgt4,8}=\frac{1}{6}(q-1)(q-2)(q-3)
\label{ccgt4_8}
\eeq
\beq
c_{cgt4,9}=\frac{1}{6}q(q-1)(q-5)
\label{ccgt4_9}
\eeq
\beq
c_{cgt4,j}=1 \quad {\rm for} \ \ j=11,12
\label{ccgt4_1112}
\eeq
\beq
c_{cgt4,j}=\frac{1}{2}q(q-3) \quad {\rm for} \ \ j=15,16 \ \ {\rm and} \ \ 
20 \le j \le 22, \ \ 27 \le j \le 30 
\label{ccgt4_15}
\eeq
\beq
c_{cgt4,j}=q-1 \quad {\rm for} \ \ 17 \le j \le 19 \ \ {\rm and} \ \ 
23 \le j \le 26 \ .
\label{ccgt4_17}
\eeq
Finally, the coloring matrix has a zero eigenvalue, 
\beq
\lambda_{cgt4,41}=0
\label{lamcgt441}
\eeq
with multiplicity
\beq
c_{cgt4,41}=\frac{1}{12}q(q-1)(3q^2-17q+40) \ . 
\label{ccgt4_41}
\eeq

It follows from the general relation (\ref{cgtritorusidentity}) that for
$L_x=0$ mod 8, this $L_y=4$ crossing-subgraph strip of the triangular lattice
reduces to the regular $L_y=4$ toroidal strip of the triangular lattice.  This
gives insight into the occurrence of the phase factors in several of the
$\lambda_{tt4,j}$ terms.  For this strip, we have $N_{cgt4,up,\lambda}=2$,
$N_{cgt4,opd,\lambda}=4$, and $N_{cgt4,ops,\lambda}=26$.  Further,
$N_{cgt4,ops,r,\lambda}=2$ and $N_{cgt4,ops,i,\lambda}=24$.  Hence, by the
general relation (\ref{nlamcgtritorus}), for $L_x=0$ mod 8, where the $L_y=4$
crossing-subgraph strip reduces to the toroidal strip of the triangular
lattice, the number of nonzero terms is reduced to
$N_{tt4,\lambda}=40-2-1=37$.  For $L_x=1$ mod 4 where the $L_y=4$
crossing-subgraph strip reduces to the Klein bottle strip of the triangular
lattice, the number of nonzero terms is reduced, according to the general
formula (\ref{nlamcgtrikb}), to $N_{tk4,\lambda}= 40-2-26=12$.  These numbers
agree with our exact calculations presented in the text.

\vfill
\eject

\begin{thebibliography}{99}

\bibitem{potts}{R. B. Potts, Proc. Camb. Phil. Soc. {\bf 48} (1952) 106.}

\bibitem{wurev}{F. Y. Wu, Rev. Mod. Phys. {\bf 54} (1982) 235.}

\bibitem{al}{M. Aizenman and E. H. Lieb, J. Stat. Phys. {\bf 24} (1981) 279.}

\bibitem{cw}{Y. Chow and F. Y. Wu, Phys. Rev. {\bf B36} (1987) 285.}

\bibitem{rrev}{R. C. Read, J. Combin. Theory {\bf 4} (1968) 52.}

\bibitem{rtrev}{R. C. Read and W. T. Tutte, ``Chromatic Polynomials'',
in {\it Selected Topics in Graph Theory, 3}, (Academic Press, New York, 1988),
p. 15.}

\bibitem{bbook}{N. L. Biggs, {\it Algebraic Graph Theory} (2nd ed., Cambridge
Univ. Press, Cambridge, 1993).}

\bibitem{w}{R. Shrock and S.-H. Tsai, Phys. Rev. {\bf E55} (1997) 5165.}

\bibitem{bkw}{S. Beraha, J. Kahane, and N. Weiss, J. Combin. Theory B
{\bf 28} (1980) 52.}

\bibitem{bds}{N. L. Biggs, R. M. Damerell, and D. A. Sands, J. Combin. Theory
B {\bf 12} (1972) 123.}

\bibitem{bm}{N. L. Biggs and G. H. Meredith, J. Combin. Theory B{\bf 20}
(1976) 5.}

\bibitem{b}{N. L. Biggs, Bull. London Math. Soc. {\bf 9} (1977) 54.}

\bibitem{readcarib81}{R. C. Read, in Proc. 3rd Caribbean Conf. on Combin. and
Computing (1981).}

\bibitem{readcarib88}{R. C. Read, in Proc. 5th Caribbean Conf. on Combin. and
Computing (1988).}

\bibitem{read91}{R. C. Read and G. F. Royle, in {\it Graph Theory,
Combinatorics, and Applications} (Wiley, NY, 1991), vol. 2, p. 1009.}

\bibitem{tk}{N. L. Biggs and R. Shrock, J. Phys. A (Letts) {\bf 32}, L489
(1999).}

\bibitem{t}{S.-C. Chang and R. Shrock, YITP-SB-99-50 (Oct. 1999), 
cond-mat/0005236.}

\bibitem{pg}{R. Shrock and S.-H. Tsai, J. Phys. A {\bf 32} (1999) 5053.}

\bibitem{wcyl}{R. Shrock and S.-H. Tsai, Phys. Rev. {\bf E60} (1999) 3512.}

\bibitem{wcy}{R. Shrock and S.-H. Tsai, Physica A {\bf 275} (2000) 429.}

\bibitem{pm}{R. Shrock, Phys. Lett. {\bf A261} (1999) 57.}

\bibitem{bcc}{R. Shrock, in the {\it Proceedings of the 1999 British
Combinatorial Conference, BCC99} (July, 1999).}

\bibitem{matmeth}{N. L. Biggs, reports LSE-CDAM-99-03,05.} 

\bibitem{k}{S.-C. Chang and R. Shrock, Phys. Rev. E, in press
(cond-mat/0004129).}

\bibitem{s4}{S.-C. Chang and R. Shrock, Physica A, in press
(cond-mat/0004161).}

\bibitem{cf}{S.-C. Chang and R. Shrock, YITP-SB-00-10, cond-mat/0005232.} 

\bibitem{strip}{M. Ro\v{c}ek, R. Shrock, and S.-H. Tsai, Physica
{\bf A252} (1998) 505.}

\bibitem{strip2}{M. Ro\v{c}ek, R. Shrock, and S.-H. Tsai, Physica {\bf A259}
(1998) 367.}

\bibitem{hs}{R. Shrock and S.-H. Tsai, Physica {\bf A259} (1998) 315.}

\bibitem{w2d}{R. Shrock and S.-H. Tsai, Phys. Rev. {\bf E58} (1998) 4332;
cond-mat/9808057.}

\bibitem{lieb}{E. H. Lieb, Phys. Rev. {\bf 162} (1967) 162.}

\bibitem{ssbounds}{J. Salas and A. Sokal, J. Stat. Phys. {\bf 86} (1997)
551.}

\bibitem{baxter}{R. J. Baxter, J. Phys. A {\bf 20} (1987) 5241.} 

\end{thebibliography}
\end{document}